\documentclass[aps,prd,preprint,a4paper,showpacs,nofootinbib,
superscriptaddress]{revtex4-1}
\usepackage{bm}
\usepackage{indentfirst}
\usepackage{amsmath}
\usepackage{graphicx}
\usepackage{float}
\usepackage{amssymb}
\usepackage{subfigure}
\usepackage{amssymb}
\usepackage{hyperref}
\usepackage{epstopdf}
\usepackage[section]{placeins}

\usepackage[utf8]{inputenc}
\hypersetup{
    colorlinks=true,
    linkcolor=red,
    citecolor=blue,
}
\usepackage{color}
\usepackage[T1]{fontenc}
\usepackage{txfonts}

\begin{document}
\baselineskip=0.5 cm

\title{Shadow and near-horizon characteristics of the acoustic charged black hole in curved spacetime}
\author{Ru Ling}
\email{rubidiumling@gmail.com}
\affiliation{Center for Gravitation and Cosmology, College of Physical Science and Technology, Yangzhou University, Yangzhou 225002, China}
\author{Hong Guo}
\email{gh710105@gmail.com}
\affiliation{Center for Gravitation and Cosmology, College of Physical Science and Technology, Yangzhou University, Yangzhou 225002, China}
\affiliation{School of Aeronautics and Astronautics, Shanghai Jiao Tong University, Shanghai 200240, China}
\author{Hang Liu}
\email{hangliu@sjtu.edu.cn}
\affiliation{Center for Gravitation and Cosmology, College of Physical Science and Technology, Yangzhou University, Yangzhou 225002, China}
\affiliation{School of Physics and Astronomy, Shanghai Jiao Tong University, Shanghai 200240, China}
\author{Xiao-Mei Kuang}
\email{xmeikuang@yzu.edu.cn}
\affiliation{Center for Gravitation and Cosmology, College of Physical Science and Technology, Yangzhou University, Yangzhou 225002, China}
\author{Bin Wang}
\email{wang_b@sjtu.edu.cn}
\affiliation{School of Aeronautics and Astronautics, Shanghai Jiao Tong University, Shanghai 200240, China}
\affiliation{Center for Gravitation and Cosmology, College of Physical Science and Technology, Yangzhou University, Yangzhou 225002, China}

\begin{abstract}
\baselineskip=0.5 cm
In this paper, we first analyze the horizon structure of the acoustic charged black hole in curved spacetime, and then study its acoustic shadow as well as the near-horizon characteristics including the quasinormal modes (QNM) frequencies and analogue Hawking radiation. We find that the radius of the acoustic shadow for acoustic charged black hole is larger than that for Reissner-Nordstr\"{o}m (RN) black hole, and  both of them are suppressed by increasing the black hole charge because their related outer horizons become smaller.
Then the QNM frequencies under scalar field perturbation and its eikonal limit are computed via numeric method and acoustic shadow, respectively. We find that the acoustic charged black hole is stable under the perturbation and the QNM frequencies are much weaker than that for the astrophysical black hole. Moreover, as the tuning parameter increases, the perturbation oscillates more mildly and its damping time becomes longer, while as the charge increases, the oscillation is enhanced slightly and the perturbation decays a little faster which is different from that in RN black hole. Finally, we numerically study the analogue Hawking radiation. We find that the grey-body factor and energy emission rate are suppressed by the angular number and the charge, but they do not monotonically depend on the tuning parameter in the acoustic charged black hole. The behavior of the energy emission rate affected by the parameters could be explained by the dependent behavior of the Hawking temperature. We expect that our results could shed light to the study of black holes in both theoretical and experimental perspectives.
\end{abstract}

\maketitle
\tableofcontents

\section{Introduction}
Astrophysical observation on black holes is one of the most significant missions in gravitational and cosmological physics. The landmark observation of GW150914 \cite{TheLIGOScientific:2016wfe} published by LIGO and Virgo Collaboration  was the first gravitational wave event from the merger of a binary system of black holes. Besides, the Event Horizon Telescope (EHT) Collaboration published  an image of the shadow of the black hole at the center of the galaxy M87 \cite{Akiyama:2019cqa,Akiyama:2019brx,Akiyama:2019sww,Akiyama:2019bqs,Akiyama:2019fyp,Akiyama:2019eap}. They are major breakthrough in gravitational physics especially for  the development of black hole physics. These two observations are the direct evidence of the existence of black holes in our universe, and they open a bright window to further study the properties of black holes. Though past decades witnessed remarkable progress on the black hole physics, more important and interesting characteristics of black holes deserve further investigation, especially its near-horizon properties. For instance, quasinormal modes (QNMs) \cite{Press:1971wr,Vishveshwara:1970zz} and Hawking radiation \cite{Hawking:1974rv,Hawking:1974sw} are important features of black holes  which could help us understand the general relativity, thermodynamics, statistics, and quantum mechanics. However, the observation of these features of black holes have encountered great challenges due to the technical limitations and theoretical accuracy requirements.

To improve the situation, a remarkable attempt is to establish analogous black holes in the laboratory which provides potential connections between astrophysical phenomena and the tabletop experiments. Unruh first proposed the acoustic black hole model  in the normal nonrelativistic fluid and studied the black hole evaporation as well \cite{Unruh:1980cg}. Later on, plenty of physical phenomena were examined to find the effective geometry and mimic some astrophysical scenarios, see for examples \cite{Visser:1997ux,Cardoso:2004fi,Chen:2004zr,Benone:2014nla,Sarkar:2017puh,Liberati:2020mdr,Vieira:2014rva,Nakano:2004ha,
Barcelo:2005fc,Berti:2004ju,Chen:2006zy} and therein.
More recent extension on the analogue Hawking radiation  has been discussed  in \cite{Anacleto:2019rfn,Balbinot:2019mei,Eskin:2019tin,Eskin:2019mqi}.
The thermodynamic-like description of the two-dimensional acoustic black hole can be seen in \cite{Zhang:2016pqx}. The particle dynamics in the acoustic spacetime was also studied in \cite{Wang:2019zqw}.
Meanwhile, from the view of experiments, Lahav constructed  a sonic black hole in  Bose-Einstein condensate system \cite{Lahav:2009wx}; Then the analogue Hawking radiation and corresponding Hawking temperature were observed \cite{deNova:2018rld,Isoard:2019buh}. Also the analogous black holes have been realized in the optical systems \cite{Steinhauer:2014dra,Drori:2018ivu,Rosenberg:2020jde} and other mechanics systems \cite{Guo:2019tmr,Bera:2020doh,Blencowe:2020ygo} experimentally.


More recently, physicists have constructed the acoustic black holes from the relativistic fluids with the starting of the Abelian Higgs model \cite{Ge:2010wx,Ge:2010eu,Anacleto:2010cr,Anacleto:2011bv,Ge:2015uaa,Ge:2019our}. Especially, Ge \emph{et al} studied the acoustic black hole in the curved geometry by considering the relativistic Gross-Pitaevskii theory and Yang-Mills (YM) theory\cite{Ge:2019our}.
They constructed the acoustic black hole in general curved spacetime. Then in \cite{Guo:2020blq} we studied the near-horizon properties of the acoustic Schwarzschild black hole, and proposed the acoustic shadow of the acoustic black hole. The study of the acoustic black hole in a curved background is more realistic and significant because the black holes in our universe could be in the bath of some kind of superfluid or just the cosmological microwave, suggesting a richer and more complex structure and properties in the near-horizon region.  Moreover, the presence of an acoustic horizon could affect the nature of the near-horizon region, which should shed light to the observations of astrophysical black holes.

In this paper, we shall extend the study of curved acoustic black hole into charged case, and focus on its near-horizon properties  which connect the observable quantities of the analogue black hole.
Firstly, we investigate the acoustic black hole shadow via analyzing the null geodesic in the acoustic charged black hole. Black hole shadow optically depends on the gravitational lensing \cite{Cunha:2018acu}. In \cite{Guo:2020blq} we firstly extended the optical shadow into the acoustic shadow in the acoustic black hole in curved spacetime, which is described by a dumb region for the listener.  Then, we compute the QNM frequencies of the acoustic charged black hole under the scalar field perturbation. We also study the frequencies in eikonal limit via its relation with the acoustic shadow. We find that the acoustic charged black hole is stable under the scalar field perturbation, though the behavior of QNM complexly  depends on the model parameters. Finally, we study the analogue Hawking radiation, and the grey-body factor and its related energy emission rate are numerically investigated, which gives more information about the near-horizon structure of the acoustic charged black hole.

The structure of the present work is organized as follows. In Sec. \ref{sec=BG}, we first briefly review the acoustic black hole in curved spacetime and then analyze the metric as well as the horizon for the acoustic charged black hole. In Sec. \ref{sec=shadow}, we study the acoustic shadow of the analogue black hole by analyzing the null geodesic. In Sec. \ref{sec=scalar}, we consider the covariant scalar field and analyze its effective potential in the background. In Secs. \ref{sec=QNM} and \ref{sec=GBfactor}, we investigate the stability by computing the QNM frequencies and its eikonal limit, and then explore the Hawking radiation via calculating the grey-body factor and  energy emission rate of the acoustic charged black hole. The last section is devoted to  our conclusions and  discussions.

\section{Setup of background}\label{sec=BG}
In this section, we shall briefly review the process of constructing the acoustic black hole in general curved spacetime. Then we will derive the metric of acoustic charged black hole and then  analyze its horizon structures.

\subsection{Review of acoustic black hole in curved spacetime}
The acoustic black hole in the general curved spacetime has been constructed by Ge \emph{et al}  starting from relativistic Gross-Pitaevskii(GP) theory\cite{Ge:2019our}. In this subsection we shall briefly review their construction and show how the curved acoustic black hole emerges from the GP theory. One could start with the action for a complex scalar field $\varphi$ as
\begin{equation}\label{eqaction-scalar}
	S=\int d^4x\sqrt{-g}(|\partial_\mu \varphi|^2+m^2 |\varphi|^2-\frac{b}{2}|\varphi|^4),
\end{equation}
where in GP theory $b$ is a constant and $m^2$ is a parameter depending on temperature via $m^2\sim (T-T_c)$. The equation of motion for $\varphi$ derived from the above action is
\begin{equation}\label{kgequ}
\Box\varphi+m^2\varphi-b|\varphi|^2\varphi=0.
\end{equation}

Considering that the scalar field  propagates in a fixed static background spacetime
\begin{equation}\label{bgmetric}
	ds_{bg}^2=g_{tt}dt^2+g_{rr}dr^2+g_{\vartheta\vartheta}d\vartheta^2+g_{\phi\phi}d\phi^2.
\end{equation}
One could further set the form of scalar field  as $\varphi=\sqrt{\rho(\vec{x},t)}e^{i\theta(\vec{x},t)}$ with $\rho=\rho_0+\epsilon\rho_1$ and  $\theta=\theta_0+\epsilon\theta_1$, where $(\rho_0,\theta_0)$ could be treated as the background solution in the fixed spacetime while  $(\rho_1,\theta_1)$ is the fluctuations.
Then the Klein-Gordon equation \eqref{kgequ} in the long-wavelength limit leads to series of equations with different orders of $\epsilon$.
Among them, the leading order is for the background scalar field
\begin{equation}\label{eqbrho}
b\rho_0=m^2-g^{\mu\nu}\partial_\mu\theta_0\partial_\nu\theta_0\equiv m^2-v_{\mu}v^{\mu}
\end{equation}
where $v_{\mu}$ are defined as $v_0=-\partial_t \theta_0$, $v_i=\partial_i \theta_0$ ($i=r,\vartheta, \phi$). The sub-leading order is a relativistic equation which governs the propagation of the phase fluctuation
\begin{equation}\label{eqFluc}
	\frac{1}{\sqrt{-\mathcal{G}}}\partial_\mu(\sqrt{-\mathcal{G}}\mathcal{G}^{\mu\nu}\partial_\nu\theta_1)=0.
\end{equation}
Subsequently, one could extract and derive an effective metric $\mathcal{G}_{\mu\nu}$ from \eqref{eqFluc}  as
\begin{equation}\label{element}
	\mathcal{G}_{\mu\nu}=\frac{c_s}{\sqrt{c^2_s-v_{\mu}v^{\mu}}}
\begin{pmatrix}
g_{tt}(c^2_s- v_i v^i) & \vdots & {-v_iv_t}\cr
\cdots\cdots\cdots\cdots & \cdot &\cdots\cdots\cdots\cdots\cdots\cdots\cr
 -v_iv_t & \vdots & {g_{ii}}(c^2_s-v_\mu v^\mu)\delta^{ij}+v_i v_j\cr
\end{pmatrix}
\end{equation}
where the speed of sound is defined as $c^2_s\equiv\frac{b\rho_0}{2}$.
It is obvious that both the background spacetime $ds_{bg}^2$ and the background four velocity of the fluid $v_\mu$ are encoded in the metric $\mathcal{G}_{\mu\nu}$. For simple cases with $v_a=0(a=\vartheta,\phi),v_t\neq0,v_r\neq0$ and $g_{tt}g_{rr}=-1$, one could reform \eqref{element} into the line element of a static, spherical symmetric acoustic black hole as
\begin{eqnarray}\label{acousticMetrc1}
	ds^2&=& c_s\sqrt{c^2_s-v_{\mu}v^{\mu}}\bigg[\frac{c^2_s-v_rv^r}{c^2_s-v_{\mu}v^{\mu}}g_{tt}
dt^2+\frac{c^2_s}{c^2_s-v_rv^r}g_{rr}dr^2+g_{\vartheta\vartheta}d\vartheta^2+g_{\phi\phi}d\phi^2\bigg],
\end{eqnarray}
where the coordinate transformation $dt\to dt-\frac{v_tv_r}{g_{tt}(c_s^2-v_rv^r)}dr$ was employed.
\subsection{Acoustic charged black hole: Metric and acoustic horizon}\label{potential_ana}
In this paper, we are interested in the acoustic charged black hole in curved spacetime. Thus we shall take account into the static spacetime background \eqref{bgmetric} as the RN black hole
\begin{eqnarray}\label{schw}
ds^2_{bg}&=&g_{tt}dt^2+g_{rr}dr^2+g_{\vartheta\vartheta}d\vartheta^2+g_{\phi\phi}d\phi^2\nonumber\\
&=&-f(r)dt^2+\frac{dr^2}{f(r)}+r^2(d\vartheta^2+\sin^2\vartheta d\phi^2),
\end{eqnarray}
where $f(r)=1-\frac{2M}{r}+\frac{Q^2}{r^2}$. Here we consider the black hole mass $M$ and charge $Q$ satisfying $M\geq Q$ where the equality indicates the extremal RN black hole. Then one could consider a vortex orbit which is falling radially from infinity  to the RN black hole {, subsequently, the RN black hole could be in the bath of the relativistic fluid. Thus, for a static observer with radial position $r$, the radial velocity of the fluid, $v_r$, should not be smaller than the escape velocity which is $\sqrt{({2M}/{r}-{Q^2}/{r^2})}$, so that the relativistic fluid could escape from the strong attraction of the background black hole. For this consideration, we define the radial component velocity as $v_r\equiv\sqrt{({2M}/{r}-{Q^2}/{r^2})\xi}$, where the tuning parameter, $\xi$,
is required to satisfy $\xi\geq 1$ such that the relativistic fluid could move safely outside the background black hole.} Moreover, following \cite{Ge:2019our}, we could also work at the critical temperature of GP theory such that  $m^2\sim (T-T_c)$ in the action \eqref{eqaction-scalar} vanishes.
Then the equation \eqref{eqbrho} reduces to $v_{\mu}v^{\mu}=-2c_s^2$. Further rescaling $v^\mu v_\mu \rightarrow {v^\mu v_\mu}/{2c^2_s}$ gives us $v_{\mu}v^{\mu}=-1$, which could be fulfilled when the time component of the velocity is $v_t=\sqrt{f(r)[1+(2M/r-Q^2/r^2)\xi]}$.
Subsequently, the static, spherical symmetric  metric of the acoustic charged black hole reduced from Eq.(\ref{acousticMetrc1}) is
\begin{eqnarray}\label{acousticMetrc2}
&&ds^2=\sqrt{3} c^2_s\bigg[-\mathcal{F}(r)dt^2+\frac{dr^2}{\mathcal{F}(r)}+r^2(d\vartheta^2+sin^2\vartheta d\phi^2)\bigg],\nonumber\\
&&\mathrm{with} ~~~ \mathcal{F}(r)=\bigg(1-\frac{2M}{r}+\frac{Q^2}{r^2}\bigg)\bigg[1- \xi \left(\frac{2 M}{r}-\frac{Q^2}{r^2}\right)\left(1-\frac{2M}{r}+\frac{Q^2}{r^2}\right)\bigg],\label{fr}
\end{eqnarray}
where the valid regime of the tuning parameter for the existence of acoustic charged black holes will
be discussed later. Note that as $\xi\to0$, the metric \eqref{fr} recovers the RN metric \eqref{schw}. While $\xi\to+\infty$, the acoustic charged black hole should fill the whole spacetime because the {escape velocity $v_r$ reaches infinity.} It also means that in this limit the event horizon of  acoustic black hole goes to infinity as we will show soon. In addition, as $Q\to0$, the radial component is $v_r\sim \sqrt{{2M\xi}/{r}}$, and then the outcome reproduces that in acoustic Schwarzschild spacetime discussed in \cite{Ge:2019our,Guo:2020blq}.

Let us then analyze the horizon structure of the whole spacetime in our setup. For convenience we shall then set $\sqrt{3} c^2_s=1$.
The vanishing of the metric function $\mathcal{F}(r)=0$ has six roots: The vanishing of the first term, $(1-\frac{2M}{r}+\frac{Q^2}{r^2})=0$ gives us the optical event horizon $r_h=M+\sqrt{M^2-Q^2}$ and Cauchy horizon $r_{c}=M-\sqrt{M^2-Q^2}$, while the vanishing of the second term, $1- \xi (\frac{2 M}{r}-\frac{Q^2}{r^2})(1-\frac{2M}{r}+\frac{Q^2}{r^2})=0$, gives us another four solutions
\begin{eqnarray}\label{solution1}
r_{a1}=\frac{M \xi}{2} - \frac{1}{2} M\Xi - \frac{1}{2}\sqrt{X_{\xi}-Y_{\xi}}, \\
r_{a2}=\frac{M \xi}{2} - \frac{1}{2} M\Xi + \frac{1}{2}\sqrt{X_{\xi}-Y_{\xi}}, \\
r_{a3}=\frac{M \xi}{2} + \frac{1}{2} M\Xi - \frac{1}{2}\sqrt{X_{\xi}+Y_{\xi}}, \\
r_{a4}=\frac{M \xi}{2} + \frac{1}{2} M\Xi + \frac{1}{2}\sqrt{X_{\xi}+Y_{\xi}},\label{solution4}
\end{eqnarray}
where $\Xi$, $X_{\xi}$ and $Y_{\xi}$ are defined, respectively, as
\begin{eqnarray}\label{Xi}
\Xi&=&\sqrt{\xi^2-4\xi},\\
X_{\xi}&=& 2 M^2 \xi ^2-4 M^2 \xi-2 \xi  Q^2, \\
Y_{\xi}&=& \frac{8 M^3 \xi ^3-8 M \xi  \left(4 M^2 \xi +\xi  Q^2\right)+32 M \xi  Q^2}{4 M \Xi}.
\end{eqnarray}
These four solutions are real only if $\xi\geq 4$, and when $\xi=4$ one has $r_{a4}=r_{a2}>r_{a1}=r_{a3}$ which could be treated as the extremal case of the acoustic charged black hole.

\begin{figure}[thbp]
\center{
\includegraphics[scale=0.51]{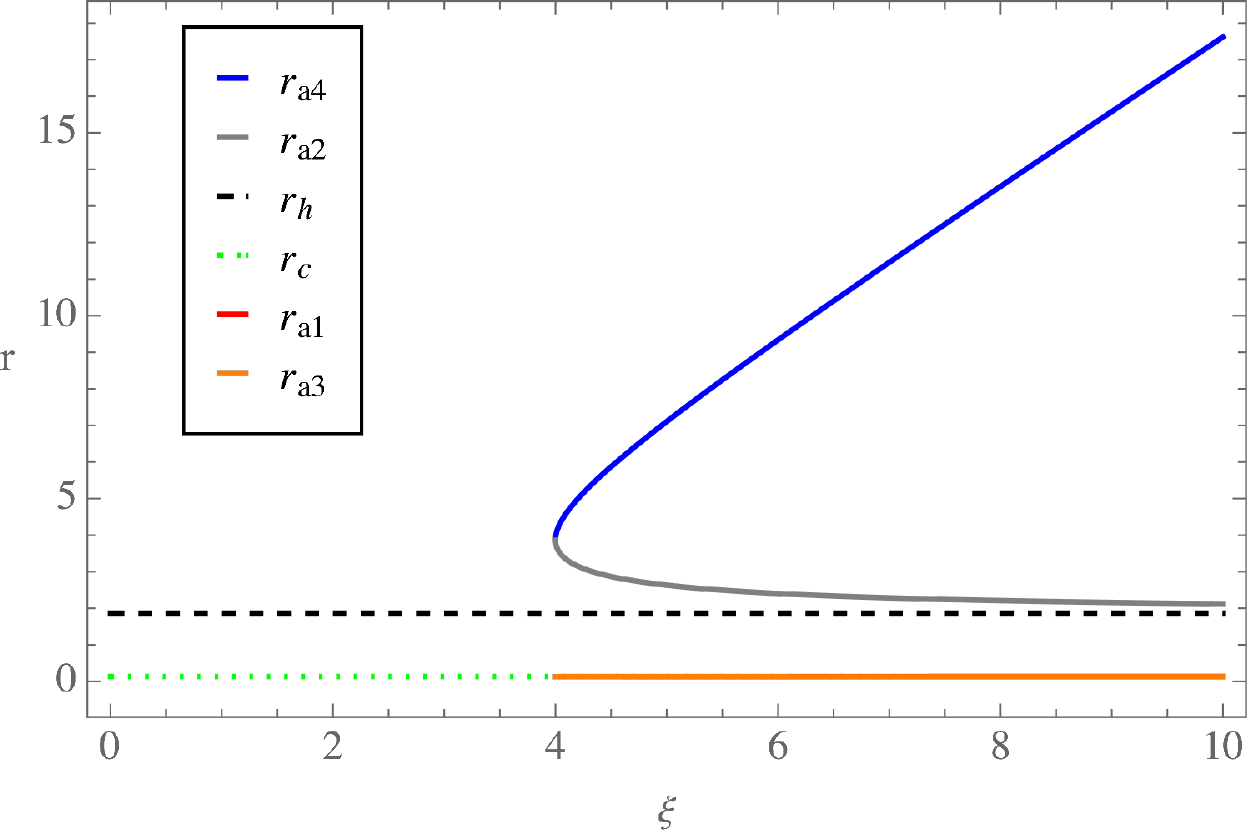}\hspace{1cm}
\includegraphics[scale=0.54]{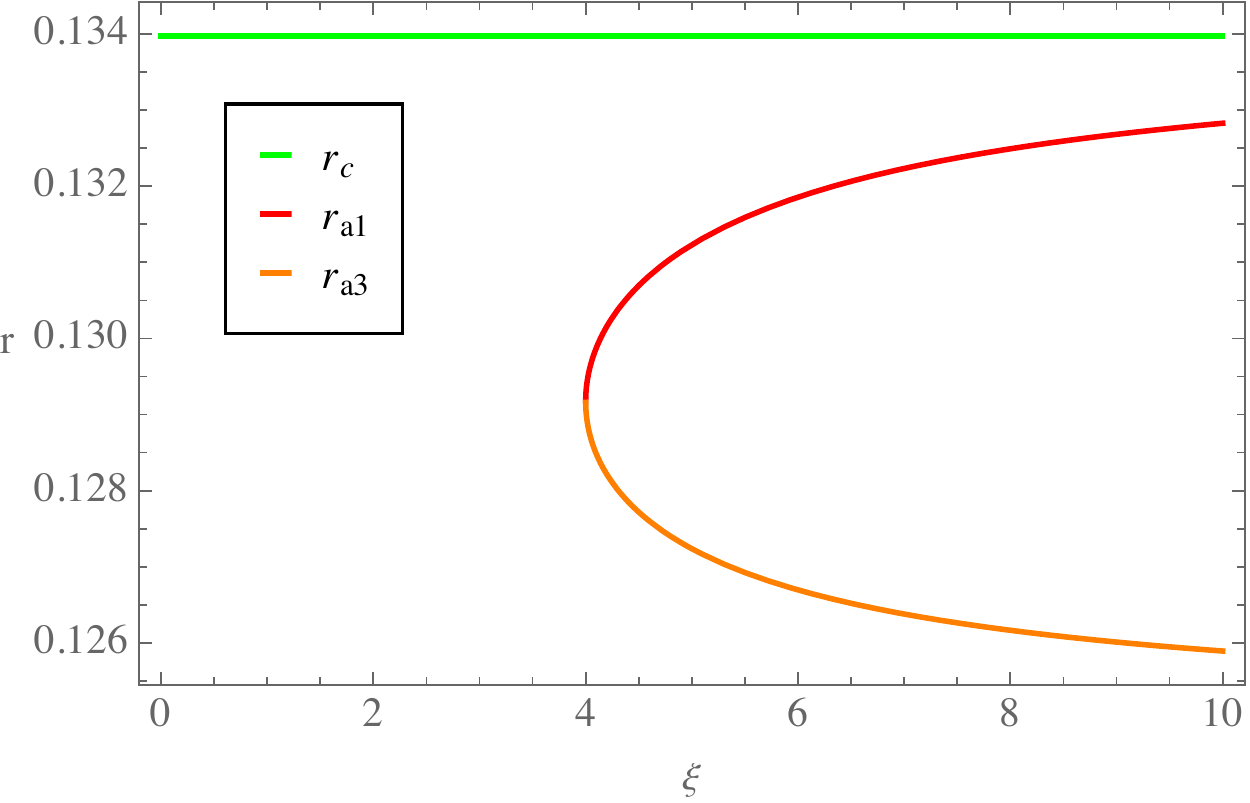}
\caption{Left: The image corresponds to event horizon $r_{h}$, Cauchy horizon $r_{c}$ and the roots Eq.(\ref{solution1}-\ref{solution4}) for the vanishing redshift function $\mathcal{F}(r)=0$ as a function of $\xi$. Right: The enlarged roots $r_{c}$, $r_{a1}$ and $r_{a3}$ as a function of $\xi$. We set $M=1$ and $Q=1/2$.}\label{horizon_xi}}
\end{figure}

\begin{figure}[thbp]
\center{
\includegraphics[scale=0.54]{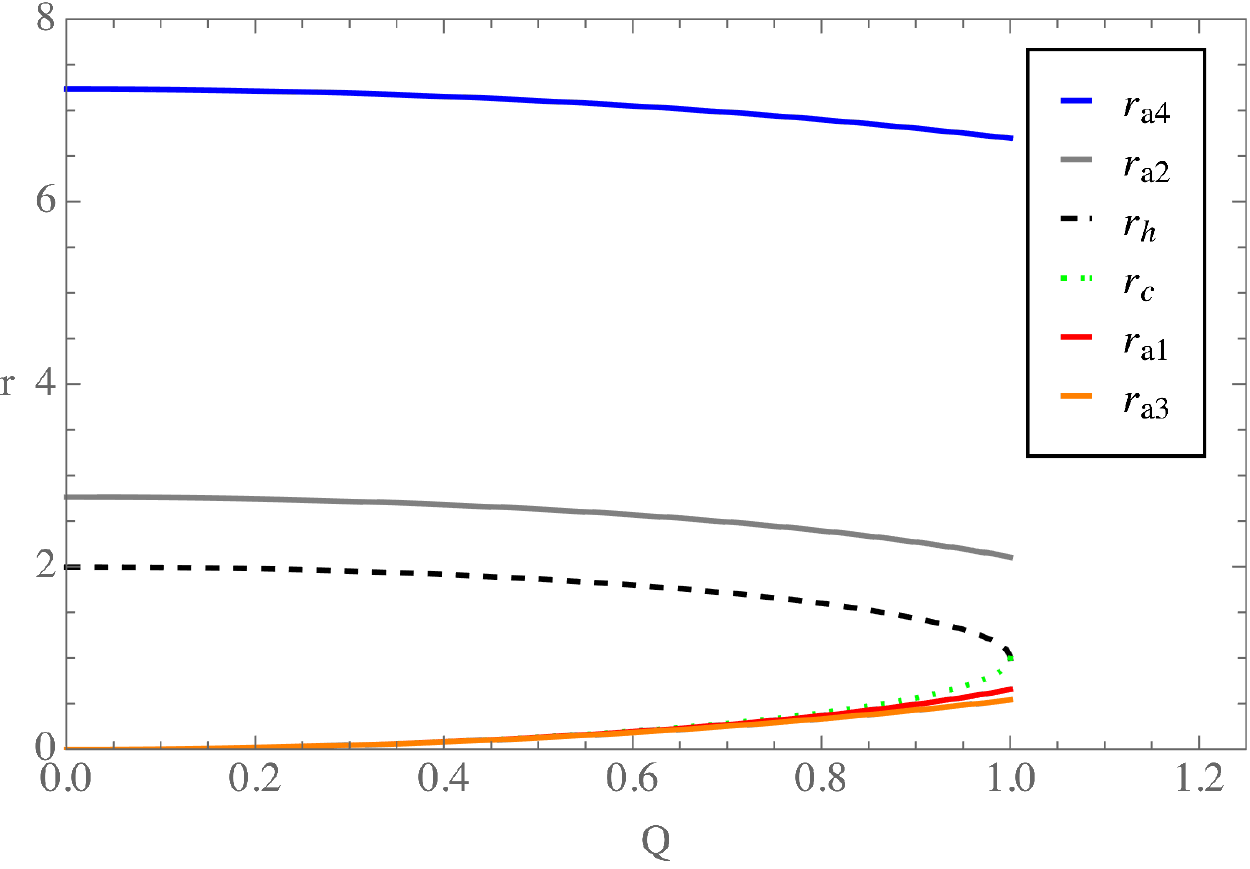}
\includegraphics[scale=0.57]{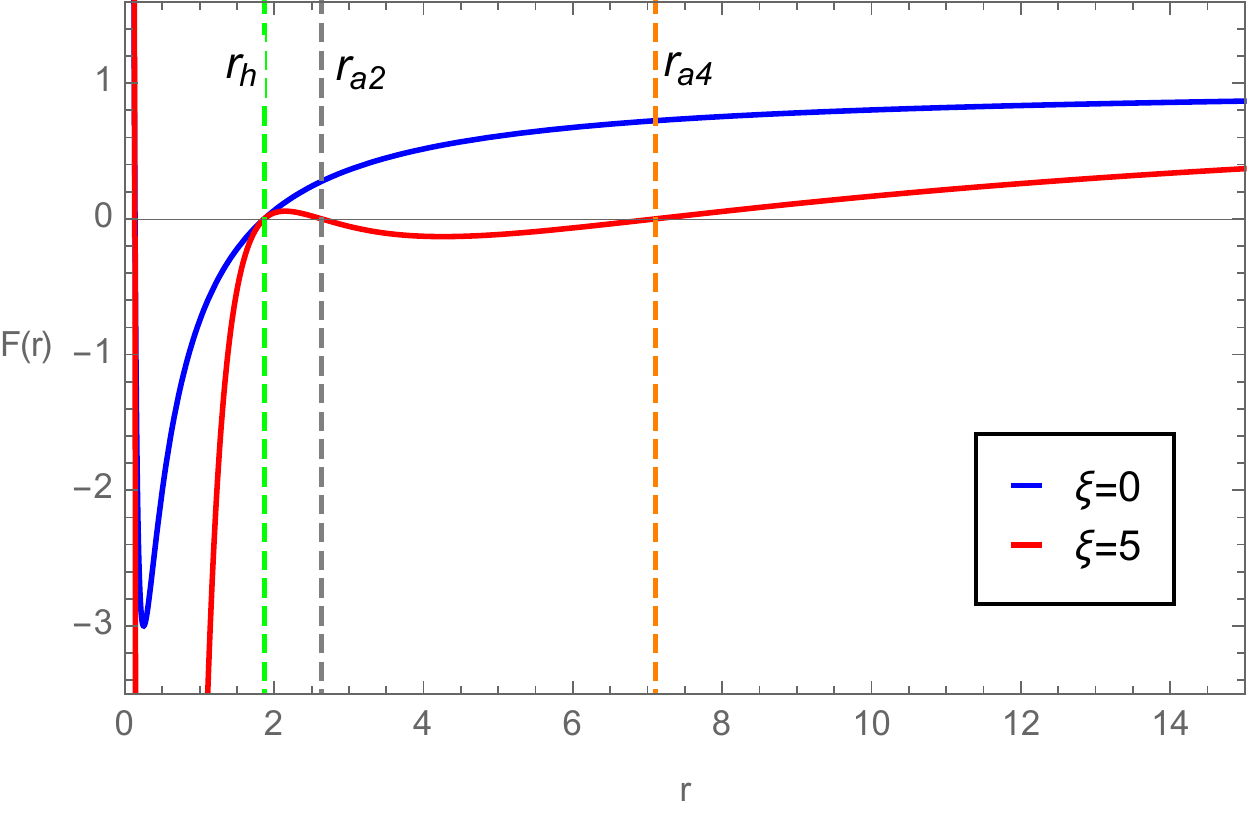}
\caption{Left: the six roots as a function of $Q$ with fixed $M=1$ and $\xi = 5$; Right: the profile of the metric function $\mathcal{F}(r)$ with $M=1, Q=1/2$. { The blue curve is for RN black hole while the red curve is for the acoustic charged black hole with $\xi=5$.} }\label{horizon_q}}
\end{figure}

We show the six roots as a function of $\xi$ in Fig. \ref{horizon_xi} with fixed $M=1$ and $Q=1/2$. In the left plot, the black dashed line represents the event horizon $r_h$ while the green dotted line represents the Cauchy horizon $r_{c}$, which are independent of $\xi$.  Obviously, the solutions $r_{a1}$ and $r_{a3}$, which are enlarged in the right plot, are always smaller than Cauchy horizon radius.
This means that these two solutions cannot exist stably because the astrophysical black hole will destroy those acoustic structures.
The radius $r_{a4}$ and $r_{a2}$ are outside the optical event horizon as expected, both of which are physical. So we could treat  $r_{a2}$ as the inner acoustic horizon while $r_{a4}$ as the outer one {which we could discuss more soon later}, and they are coincident in the extremal case with $\xi=4$. However they behave differently as the tuning parameter increases, namely, when $\xi$ increases, the inner acoustic horizon decreases and converges to the Cauchy horizon as $\xi\to+\infty$, while the outer acoustic horizon increases and finally goes to infinity, i.e, $r_{a4}\to +\infty$ as $\xi\to+\infty$. This suggests that as $\xi\to+\infty$, the sound wave could not escape from the whole spacetime as we mentioned previously. Moreover, the dependence of various solutions on the charge is shown in {the left plot of} Fig. \ref{horizon_q}. When $Q=0$, the model reduces to the acoustic Schwarzschild black hole. As $Q$ becomes non-zero, the Cauchy horizon $r_{c}$ emerges and grows to $r_h$ in the extreme case $Q=M=1$ as we all know. $r_{a1}$ and  $r_{a3}$ increases as $Q$ increases but they are always smaller than the Cauchy horizon and  not  physical. Similar as the optical event horizon $r_h$, both $r_{a2}$ and $r_{a4}$ also decrease as $Q$ increases and   $r_{a4}>r_{a2}>r_h$ always holds.


{It is worthwhile to point out that the existence of the inner acoustic horizon, $r_{a2}$, and outer acoustic horizon, $r_{a4}$, indeed exhibit colorful structure of the acoustic charged black hole. By definition, the case when the sound velocity is equal to the fluid velocity describes the acoustic horizon \cite{Visser:1997ux,Barcelo:2005fc}, indicating that the metric function $\mathcal{F}(r)$ vanishes. Then the positive value of $\mathcal{F}(r)$ implies that the speed of sound exceeds the fluid velocity, so that the sound waves could travel freely in this region, while for the region with negative  $\mathcal{F}(r)$, the sound waves will be trapped and cannot be detected by the outside world. In the right plot of Fig. \ref{horizon_q}, we show the profile of the metric function. It is interesting to note that $r_{a2}$ could be treated  as the horizon of the acoustic white hole for the observer located in the region $r_h<r<r_{a2}$. Specifically, since the fluid flow is moving from the supersonic region ($r_{a2}<r<r_{a4}$), cross the horizon $r_{a2}$, and then enter the subsonic region ($r_h<r<r_{a2}$) where the observer is located. As a consequence, the sound waves in  the subsonic region cannot propagate against the fluid flow into the supersonic region, while the other way around is allowed, and so it is reasonable to regard $r_{a2}$ as the horizon of the acoustic white hole for this observer.}

{In addition, the metric function in the right plot of Fig. \ref{horizon_q} also behaves differently in the vicinities of $r_{a2}$ and $r_{a4}$. In the vicinity of $r<r_{a2}$, $\mathcal{F}(r)>0$ while $\mathcal{F}(r)<0$  in the vicinity of $r>r_{a2}$, indicating that the inner acoustic horizon shares certain properties with the Cauchy horizon of real black hole. This similarity could suggest that more analogue properties of inner acoustic horizon deserve further study, for instance its stability. On a contrary, the metric function in the vicinity of the outer acoustic horizon behaves as $\mathcal{F}(r)>0$ for $r>r_{a4}$, while $\mathcal{F}(r)<0$ for $r<r_{a4}$, which is like the feature of the event horizon of the real black hole.}

Thus, {from all above considerations, we conclude that} in the parameter region $\xi\geq4$, the acoustic charged black hole is constructed and the spacetime thereby could be divided into four regions: Region I with  $r<r_h$ is inside the RN black hole where neither the light nor the sound wave can escape; Region II with $r_{h}<r<r_{a2}$ where light can escape {and the sound waves can also escape but cannot be detected by the observer outside this region}; Region III with $r_{a2}<r<r_{a4}$ where light can escape but the sound wave cannot; Region IV with $r>r_{a4}$ where both of the light and the sound wave can escape. {Moreover, it is also reasonable to consider the outer horizon as the acoustic horizon, i.e., $r_{ac}=r_{a4}$. Then in the following, we shall investigate various characteristics near this acoustic horizon, including the circle null geodesic, QNM frequencies, the grey-body factor and their connections.  We shall fix $M=1$ without loss of generality.}

\section{Acoustic black hole shadow}\label{sec=shadow}

Black hole shadow is one of the fingerprints of the geometry around the black hole horizon. It describes the black hole properties which depend on the gravitational lensing of the nearby radiation. Readers can see \cite{Cunha:2018acu,Perlick:2021aok} as nice reviews. Moreover, the Event Horizon Telescope group detected the black hole images with the use of the shadow properties \cite{Akiyama:2019cqa,Akiyama:2019fyp,Akiyama:2019eap} and attracted plenty of attention.
As a first attempt, in \cite{Guo:2020blq} we have studied the acoustic shadow of the curved acoustic black hole.
Theoretically, the acoustic shadow is a region of the listener's sky that is left dumb, if there are
sonic sources distributed everywhere but not between the listener and the acoustic black hole.
Acoustic shadow  describes the near-acoustic-horizon properties of the sound waves. Thus, from the above analysis of horizon structure in the acoustic charged black hole, there could exist the optical shadow around the event horizon describing  the visual boundary that the light cannot escape from the event horizon by viewers, and also exists the acoustic shadow which describes the audible boundary of the sound waves detected by static listeners. The former for the RN black hole has been studied in \cite{Zakharov:2011zz,Zakharov:2014lqa}. Thus, in this section we will focus on the acoustic shadow by analyzing the null geodesic in the acoustic charged black hole.

The geodesic motion of a null particle is governed by the Hamiltonian
\begin{equation}
	H=\frac{1}{2}\mathcal{G}_{\mu\nu}p^{\mu}p^{\nu}=0,
\end{equation}
where the metric is described in \eqref{acousticMetrc2}.
The null geodesic gives two conserved quantities,
\begin{equation}
	E=-p_t,\ \ L=p_{\phi},
\end{equation}
and  the orbit equation is reduced as
\begin{equation}
	\left(\frac{dr}{d\phi}\right)^2=V_{eff},
\end{equation}
where the effective potential reads
\begin{equation}
	V_{eff}=r^4\left(\frac{E^2}{L^2}-\frac{\mathcal{F}(r)}{r^2}\right).
\end{equation}
The circular null geodesic describes the ``acoustic sphere" when the conditions $V_{eff}=0$ and $V'_{eff}=0$ are fulfilled. Then it is easy to derive that  the radius of ``acoustic sphere" $r_{ah}$ is determined by the equation
\begin{equation}\label{eqrah}
	\frac{dh^2(r)}{dr}=0
\end{equation}
where $h(r)=\sqrt{r^2/\mathcal{F}(r)}$.  Then for a distant static listener locating at $r_L$, the detected  radius of the acoustic shadow could be defined as \cite{Perlick:2015vta}
\begin{equation}
	r_{sh}=\frac{h(r_{ah})r_L}{h(r_L)}.
\end{equation}

\begin{figure}[thbp]
\center{
\includegraphics[scale=0.35]{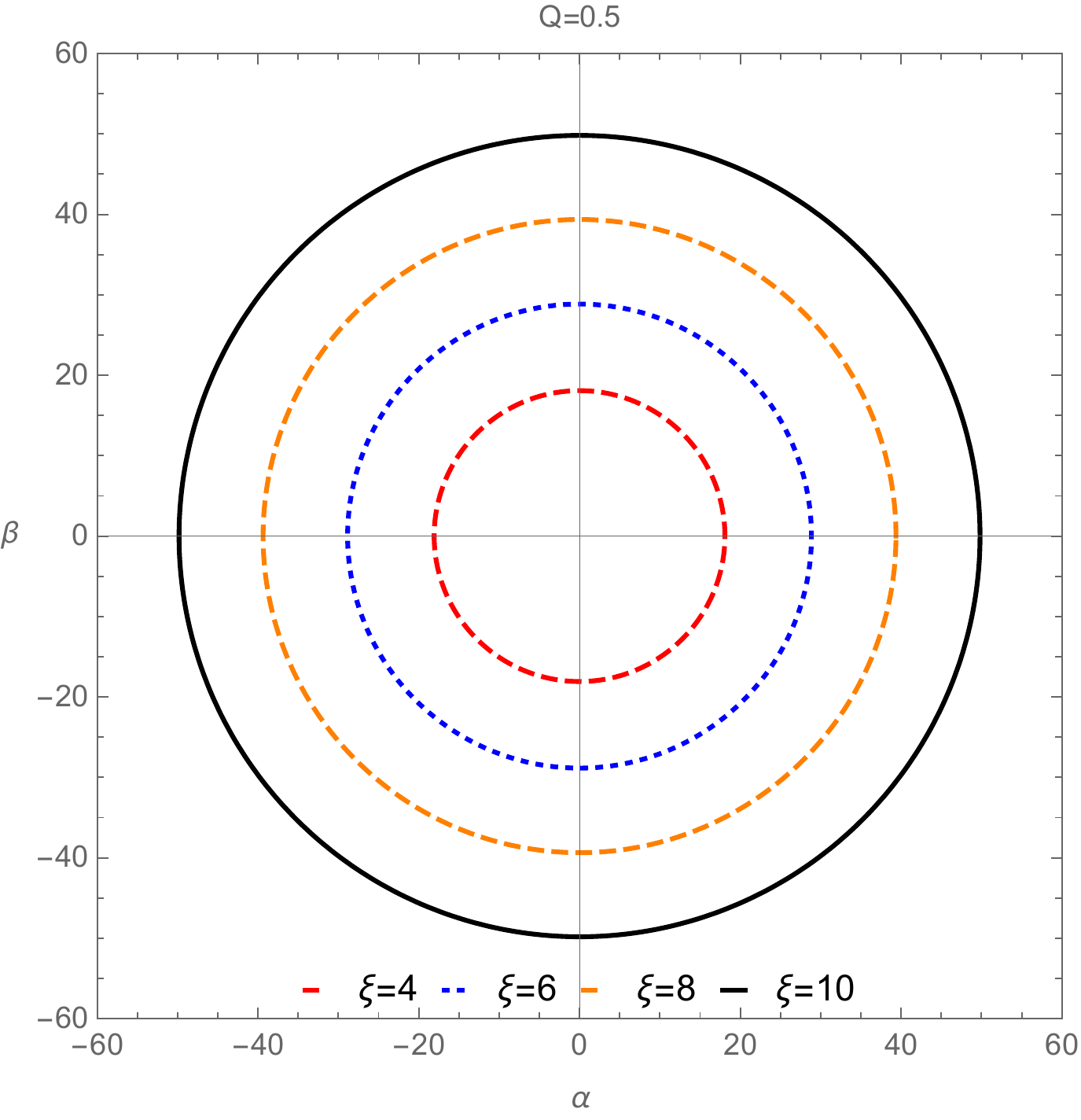}\hspace{1cm}
\includegraphics[scale=0.35]{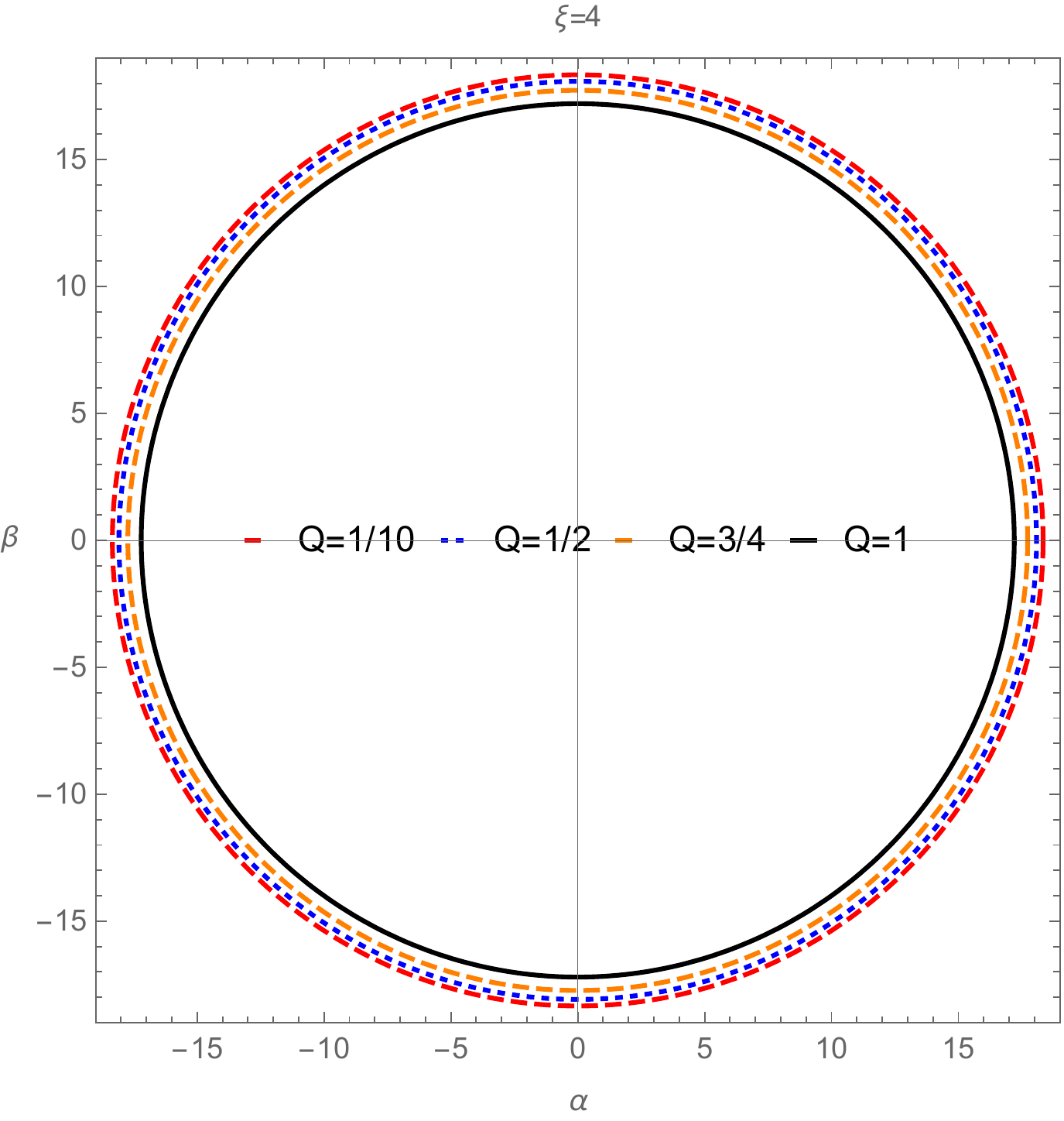}
\caption{The shadow image with different tuning parameters $\xi$ (left panel with $Q=1/2$) and black hole charge $Q$ (right panel  with $\xi=4$).}\label{shadow1}}
\end{figure}

Assuming the static listener is located at infinity far away from the acoustic black hole, we have the relation $\frac{r_L}{h(r_L)}\approx 1$ such that $r_{sh}=h(r_{ah})$. It means that once we have $r_{ah}$ in hand after solving the equation \eqref{eqrah}, we can evaluate the acoustic shadow for a distance listener.  Since the analogue black hole we considered has spherical symmetry, we shall set $r_{sh}=\sqrt{\alpha^2+\beta^2}$, and draw the acoustic shadow image in the $(\alpha,\beta)$ plane, which is shown in Fig. \ref{shadow1}. In the left plot, with fixed $Q=1/2$, the radius of the acoustic shadow is enhanced dramatically as we increase the tuning parameters $\xi$. Meanwhile, in the right plot, with fixed $\xi=4$,  the radius of the acoustic shadow becomes smaller as we increase the charge $Q$ which is similar as the optical shadow for RN black hole observed in \cite{Zakharov:2011zz,Zakharov:2014lqa}.

\begin{figure}[thbp]
\center{
\includegraphics[scale=0.4]{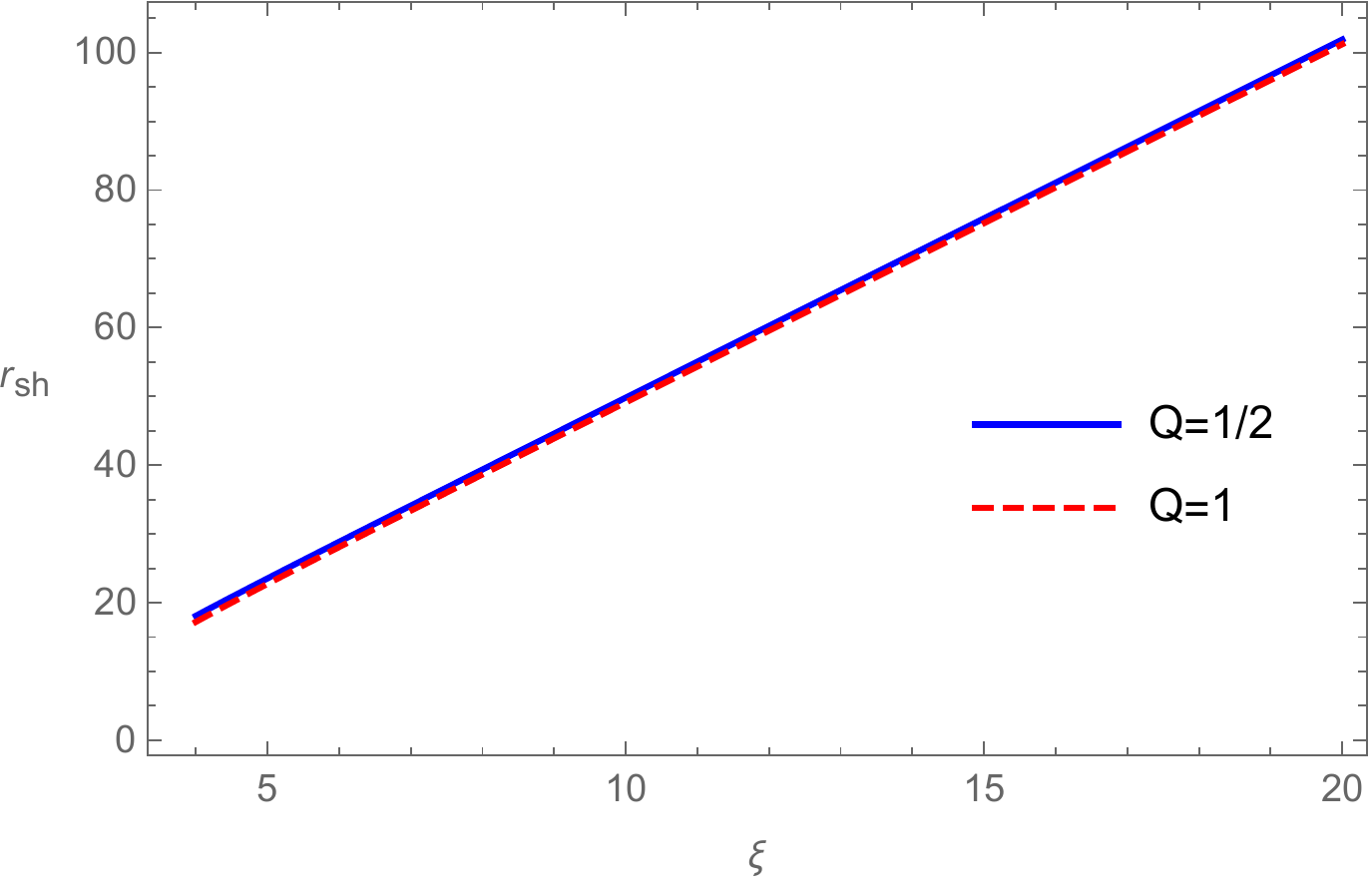}\hspace{1cm}
\includegraphics[scale=0.4]{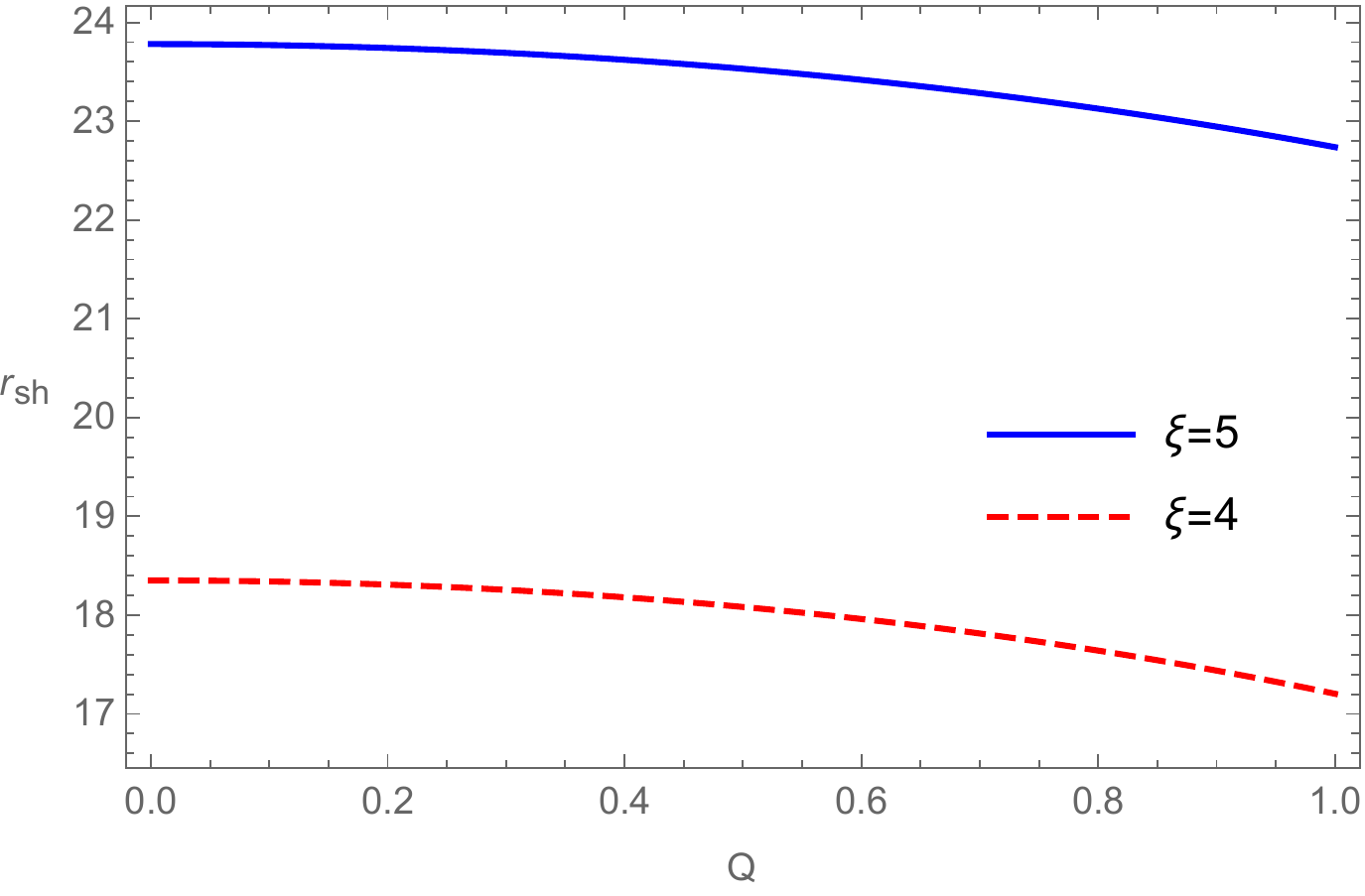}
\caption{{Left: the shadow radius $r_{sh}$ as the function of the tuning parameter $\xi$. The blue solid line is for fixed $Q=1/2$ while the red dashed line is  for the extreme case (i.e.,$r_h=r_c$) with $Q=1$. Right: the shadow radius $r_{sh}$ as the function of the charge $Q$. The blue solid line is for fixed $\xi=5$ while the red dashed line is  for the extreme case (i.e.,$r_{a2}=r_{a4}$) with $\xi=4$.}}\label{shadow}}
\end{figure}

The behavior of the acoustic shadow is reasonable according to the behavior of the radius of acoustic sphere shown in Fig. \ref{shadow}. In the left plot, the radius of the acoustic sphere increases as $\xi$ increases  due to the increasing of the acoustic horizon. This behavior was also observed in neutral case\cite{Guo:2020blq}.  In the right plot, the radius of the acoustic sphere decays slightly with the increasing of $Q$. So the acoustic shadow is the smallest when the acoustic charged black hole is in double extremal case with $Q=1$ and  $\xi=4$.



\section{Covariant scalar field equation and the effective potential }\label{sec=scalar}
In the above discussions, besides clarifying  the location of the acoustic horizon and its structure in charged background, we also analyze the shadow image of the analogue black hole. In order to understand more basic characteristics, such as QNM frequencies, grey-body factor and energy emission in the analogue Hawking radiation, we consider a test scalar field to probe the near-horizon geometrical structure. The covariant equation of the massless scalar field is
\begin{equation}
 \frac{1}{\sqrt{-\mathcal{G}}}\partial_\mu(\sqrt{-\mathcal{G}}\mathcal{G}^{\mu\nu}\partial_\nu\psi(t,r,\theta))=0,
\end{equation}
where $\mathcal{G}_{\mu\nu}$ denotes the metric components in \eqref{acousticMetrc2}. With the formula
\begin{equation}
    \psi(t,r,\theta)=\sum_{lm}e^{-i\omega t}\frac{\Psi(r)}{r}Y_{lm}(\theta),
\end{equation}
the covariant equation reduces to a Schrodinger-like equation
\begin{equation}
	\frac{d^2\Psi}{dr^2_*}+(\omega^2-V(r))\Psi=0\label{eq1}
\end{equation}
in the tortoise coordinate $r_*=\int 1/\mathcal{F}(r) dr $, and the effective potential is
\begin{equation}\label{effpoten}
V(r)= \mathcal{F}(r)\bigg[\frac{l(l+1)}{r^2}+\frac{\mathcal{F}'(r)}{r}\bigg].
\end{equation}

\begin{figure}[thbp]
\center{
\includegraphics[scale=0.42]{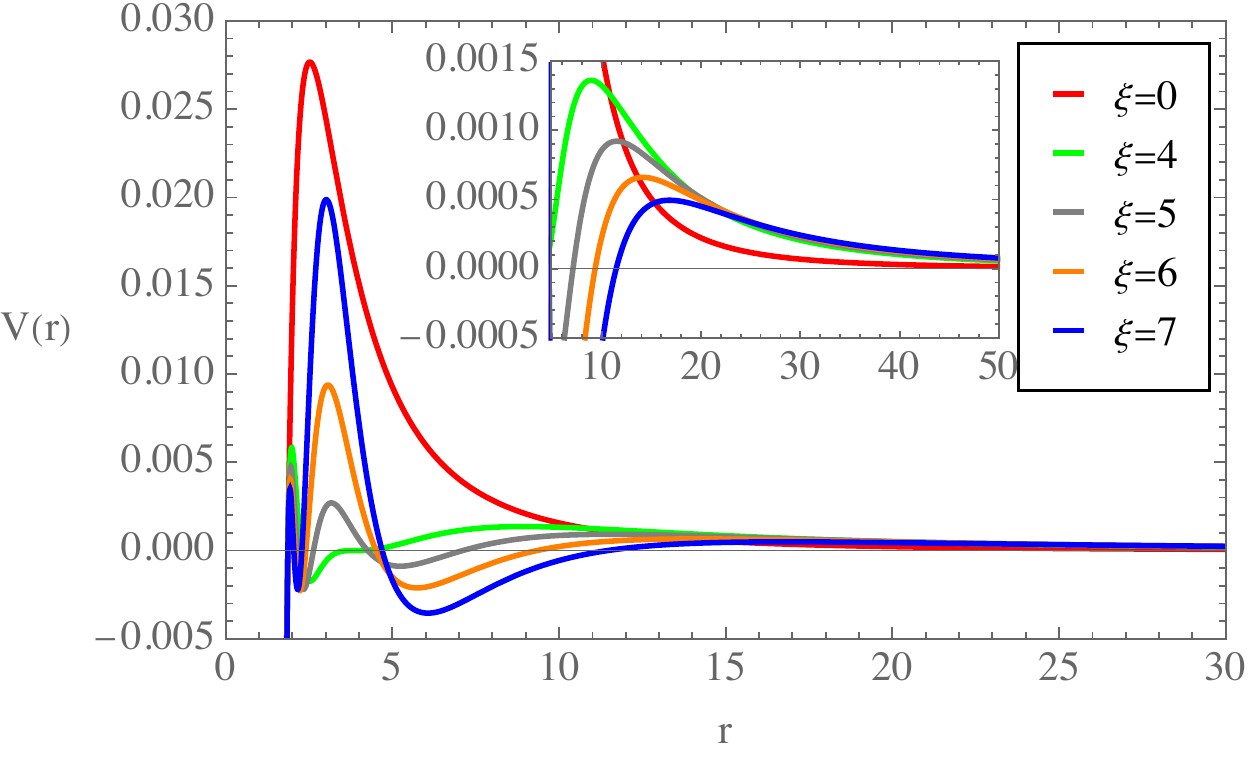}
\includegraphics[scale=0.43]{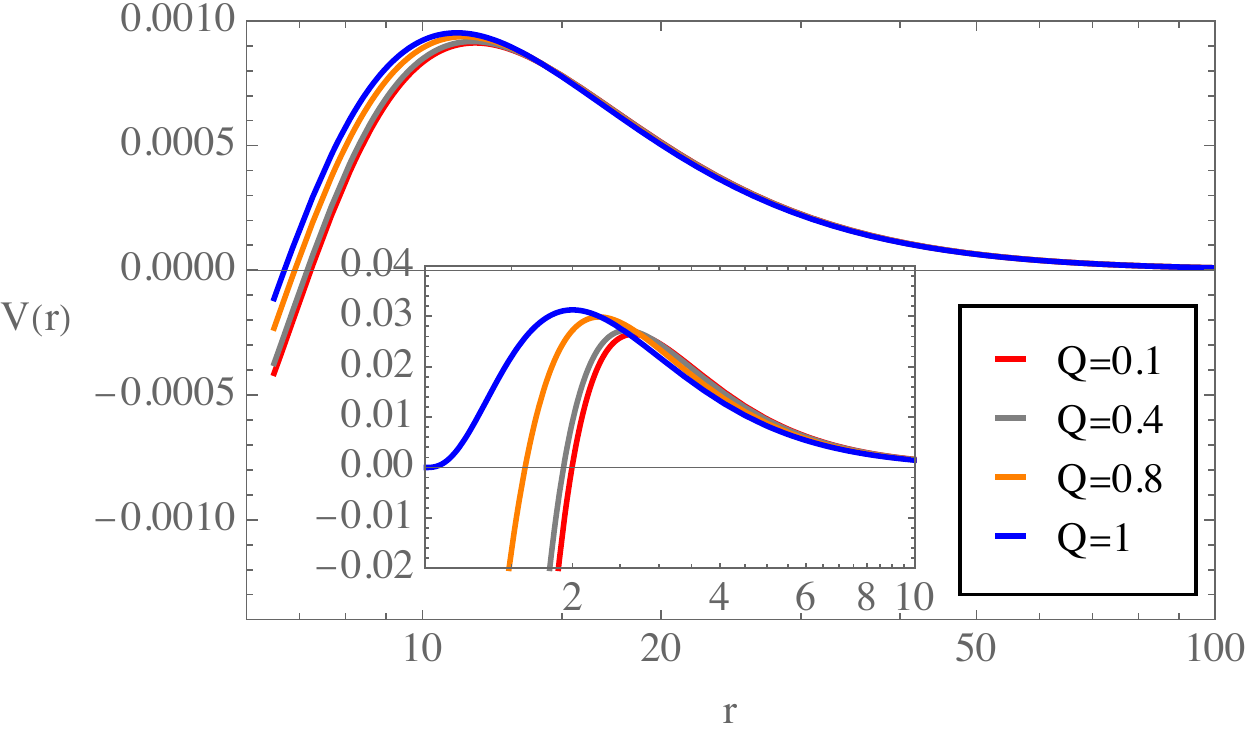}
\includegraphics[scale=0.43]{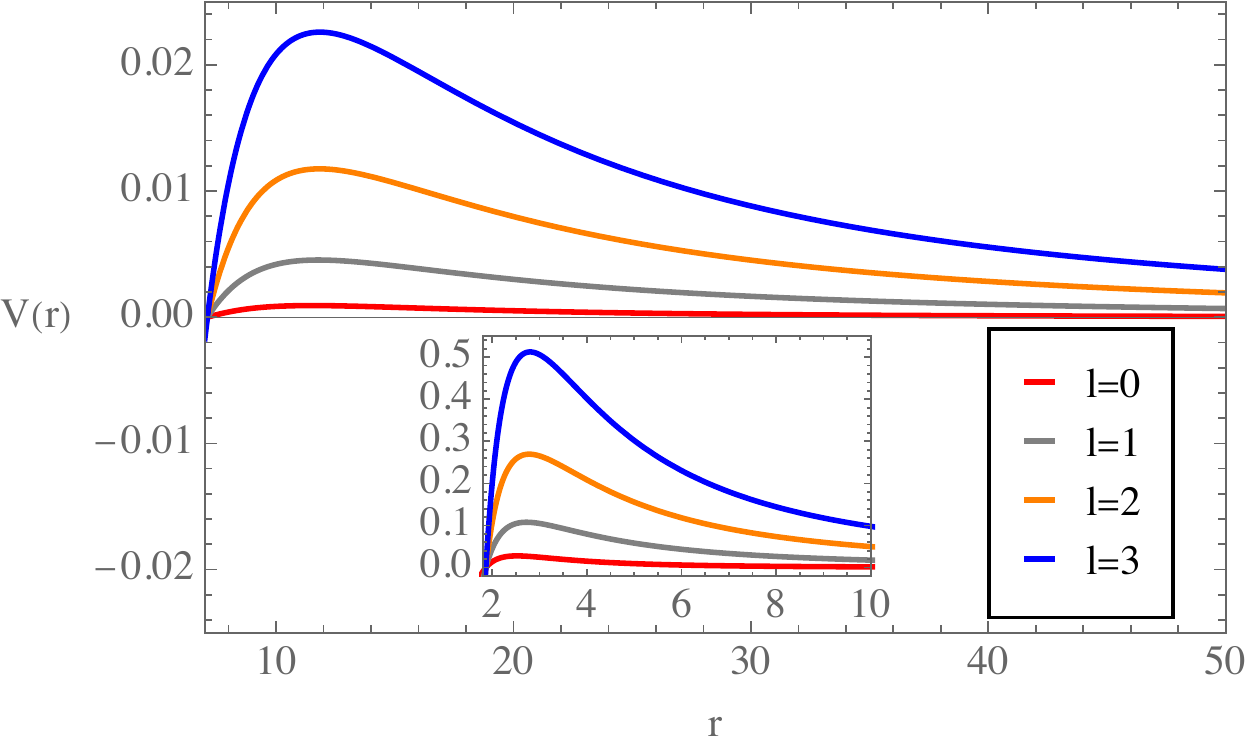}
\caption{Left: the effective potential for different $\xi$ with fixed $l=0$ and $Q=1/2$. Middle: the effective potential for different $Q$ with fixed $l=0$ and $\xi=5$. {The inset shows the potential for RN black hole with the same parameters.} Right: the effective potential for different $l$ for fixed $\xi=5$ and $Q=1/2$, {and the potential for RN black hole with the same parameters is also shown in the inset.}}\label{potential}}
\end{figure}
The behaviors of the effective potential \eqref{effpoten} are shown in Fig. \ref{potential}. {The main plots in all cases show that} as $r$  approaches to the near-horizon region from infinity, the effective potential presents a barrier and then decays rapidly to zero at the acoustic horizon $r_{ac}$. The dependence of $V(r)$ on the parameters of analogue black hole are also obvious. {Note that here we could mainly focus on the effect of the model parameters on the effective potential near the acoustic horizon, but we also reproduce the effective potentials near the event horizon for RN black hole with the same parameters (see the red line in the left plot and the insets in the middle and right plots) and do comparison.} In detail, in the left plot with fixed $l=0$ and $Q=1/2$, {the potential for the RN black hole (red curve with $\xi=0$) shows one barrier near the event horizon, while for the acoustic black holes with $\xi\geq4$, the potential barrier emerges both near the event horizon and the acoustic horizon. In addition, the barrier near the event horizon is higher than that near the acoustic horizon, which is magnified in the inset. Moreover,} as $\xi$ increases, both the acoustic horizon and the peak of the potential move to larger $r$ and the peak also becomes gentler. In the middle plot with fixed $l=0$ and $\xi=5$, as $Q$ increases, both the zero potential and barrier {near the acoustic horizon} locate at smaller $r$. This is because the location of the zero potential represents the acoustic horizon which decreases slightly as $Q$ increases as shown in Fig. \ref{horizon_q}. {This $Q-$dependent rule is similar to that for the effective potential in RN black hole, though whose barrier is higher as shown in the inset.}  With fixed $Q=1/2$ and $\xi=5$ in the right plot, we also show the profile of the potential with $l$. For larger $l$, the potential barrier {near the acoustic horizon is higher, which is also consistent with the rule of the potential for RN black hole shown in the inset.}  All these behaviors of the effective potential near the acoustic horizon with different parameters would be reflected by the near-acoustic horizon characteristics as we will discuss soon.

\section{Quasinormal mode frequencies}\label{sec=QNM}
In this section, we will study the QNM frequencies of the acoustic charged black holes. We shall first compute the QNM frequencies in numerical way. We  mainly employ the semi-analytical WKB method \cite{Konoplya:2019hlu}, and then use the asymptotic iteration method (AIM) \cite{Cho:2009cj} to verify the accuracy of the calculations. Then we shall investigate the QNM frequencies in the eikonal limit in terms of black hole shadow with the proposal in \cite{Cardoso:2008bp}.

\subsection{The numerical result}
To solve the master equation of the scalar field Eq. \eqref{eq1}, we choose  the following boundary conditions
\begin{equation}
	\Psi \sim e^{\pm i\omega_{qnm} r_*}, r_*\rightarrow \pm \infty,
\end{equation}
which means that the wave is incoming at the acoustic horizon while it is outgoing  at infinity.

Firstly we study the QNM frequencies for small $\xi$ with different angular numbers $l$ and overtone numbers $n$ as samples when we set $Q=1/2$. Table \ref{table1} shows the data for $n=0,l=0$ and $n=0,l=1$, and Table \ref{table2} shows the results for $n=1,l=1$ and $n=1,l=2$, respectively.
\begin{table}[!htbp]
\centering
\begin{tabular}{|c|c|c|c|c|}\hline
\multicolumn{1}{|c}{$\omega_{qnm}$}&\multicolumn{2}{|c|}{($n=0,\ l=0$)}& \multicolumn{2}{c|}{($n=0,\ l=1$)}\\\hline
\multicolumn{1}{|c|}{$\xi$}& WKB & AIM & WKB & AIM \\\hline
0 & 0.115596-0.105813i& - & 0.306561-0.098799i &-\\\hline
4 & 0.028671-0.019075i& - & 0.083298-0.017409i & - \\\hline
5 & 0.023608-0.016584i&0.0236989-0.0166143i&0.064572-0.015934i&0.0645715-0.0159340i\\\hline
6 & 0.019679-0.014728i&0.0196933-0.0147308i&0.052768-0.014044i&0.0527682-0.0140436i\\\hline
7 & 0.016784-0.013118i&0.0167651-0.0130978i&0.044647-0.012421i&0.0446466-0.0124199i\\\hline
8 & 0.014758-0.011704i&0.0145748-0.0117209i&0.038707-0.011089i&0.0387065-0.0110885i\\\hline
9 & 0.012895-0.010669i&0.0128813-0.0105895i&0.034169-0.009997i&0.0341686-0.0099966i\\\hline
10& 0.011531-0.009722i&0.0115443-0.0096513i&0.030587-0.009092i&0.0305868-0.0090916i\\\hline
\end{tabular}
\caption{ The QNM frequency of acoustic charged black hole with the mode $l=n=0$ and $n=0,l=1$. We set $Q=1/2$.\label{table1}}
\end{table}
\begin{table}[!htbp]
\centering
\begin{tabular}{|c|c|c|c|c|}\hline
\multicolumn{1}{|c}{$\omega_{qnm}$}&\multicolumn{2}{|c|}{($n=1,\ l=1$)}& \multicolumn{2}{c|}{($n=1,\ l=2$)}\\\hline
\multicolumn{1}{|c|}{$\xi$}& WKB & AIM & WKB & AIM \\\hline
0 & 0.279599-0.308859i& - &0.487308-0.298818i& -\\\hline
4 & 0.077742-0.053456i& - &0.135049-0.052147i& - \\\hline
5 & 0.061333-0.048742i&0.0613362-0.0487346i&0.104644-0.047929i&0.1046442-0.0479291i\\\hline
6 & 0.049943-0.043194i&0.0499384-0.0431891i&0.085225-0.042314i&0.0852247-0.0423140i\\\hline
7 & 0.042035-0.038331i&0.0420429-0.0383265i&0.071908-0.037455i&0.0719082-0.0374548i\\\hline
8 & 0.036277-0.034302i&0.0362887-0.0343130i&0.062207-0.033459i&0.0622076-0.0334588i\\\hline
9 & 0.031902-0.030978i&0.0319051-0.0309759i&0.054822-0.030176i&0.0548223-0.0301761i\\\hline
10& 0.028468-0.028212i&0.0284683-0.0282138i&0.049009-0.027453i&0.0490095-0.0274526i\\\hline
\end{tabular}
\caption{ The QNM frequency of acoustic charged black hole with the mode $n=1,l=1$ and $n=1,l=2$. We set $Q=1/2$.\label{table2}}
\end{table}

As can be seen from the tables, the real part of QNM $Re (\omega_{qnm})$ is positive and the imaginary part $Im (\omega_{qnm})$ is negative, which means that the acoustic charged black hole is in a stable state under the perturbation for small tuning parameters. Comparing to  RN black hole with $\xi=0$, the acoustic charged  black hole has much smaller magnitudes of the QNM frequency. This indicates that the signal of the QNM frequency of the acoustic black hole is much weaker than the astrophysical black hole. It means that comparing to astrophysical black hole, the QNMs of acoustic black holes oscillate more slowly and also have slower decay rate which makes it more likely to be detected.

In addition, as $\xi$ increases, $Re (\omega_{qnm})$ attenuates indicating that the strength of the oscillations  is suppressed, while the magnitudes of $Im (\omega_{qnm})$ decreases meaning that the decay of the scalar field  becomes slower.
This behavior could be attributed to the suppression of the effective potential by larger $\xi$ (see Fig. \ref{potential}). We need to point out that in both tables, the magnitudes of $Im (\omega_{qnm})$  decays monotonically. It would be necessary to compute the QNMs for larger $\xi$ to further check if it crosses the horizonal axis or not. That is to say, we have to further check the (in)stability of the acoustic charged black hole with large $\xi$.

\begin{figure}[thbp]
\center{
\includegraphics[scale=0.5]{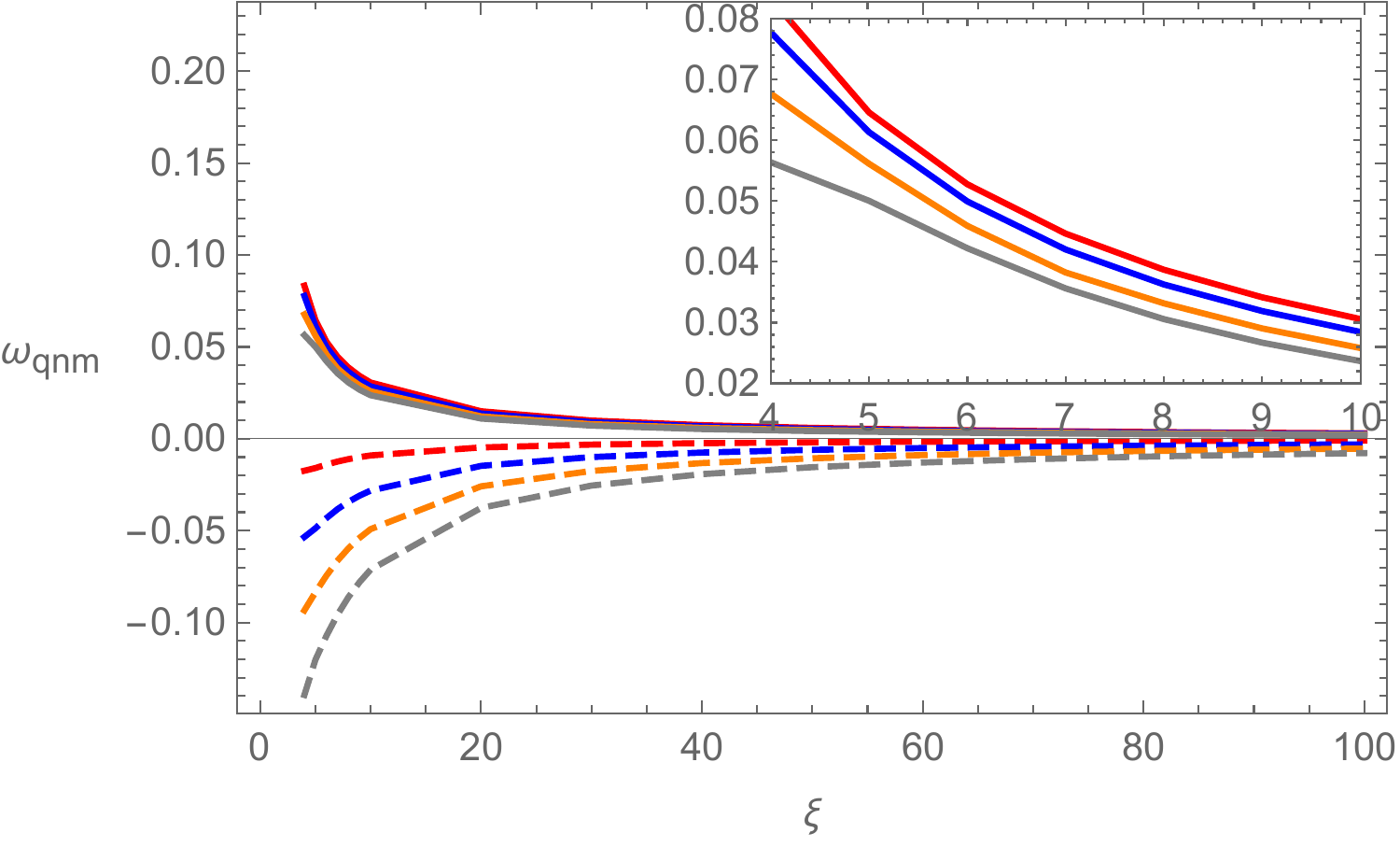}\hspace{1cm}
\includegraphics[scale=0.5]{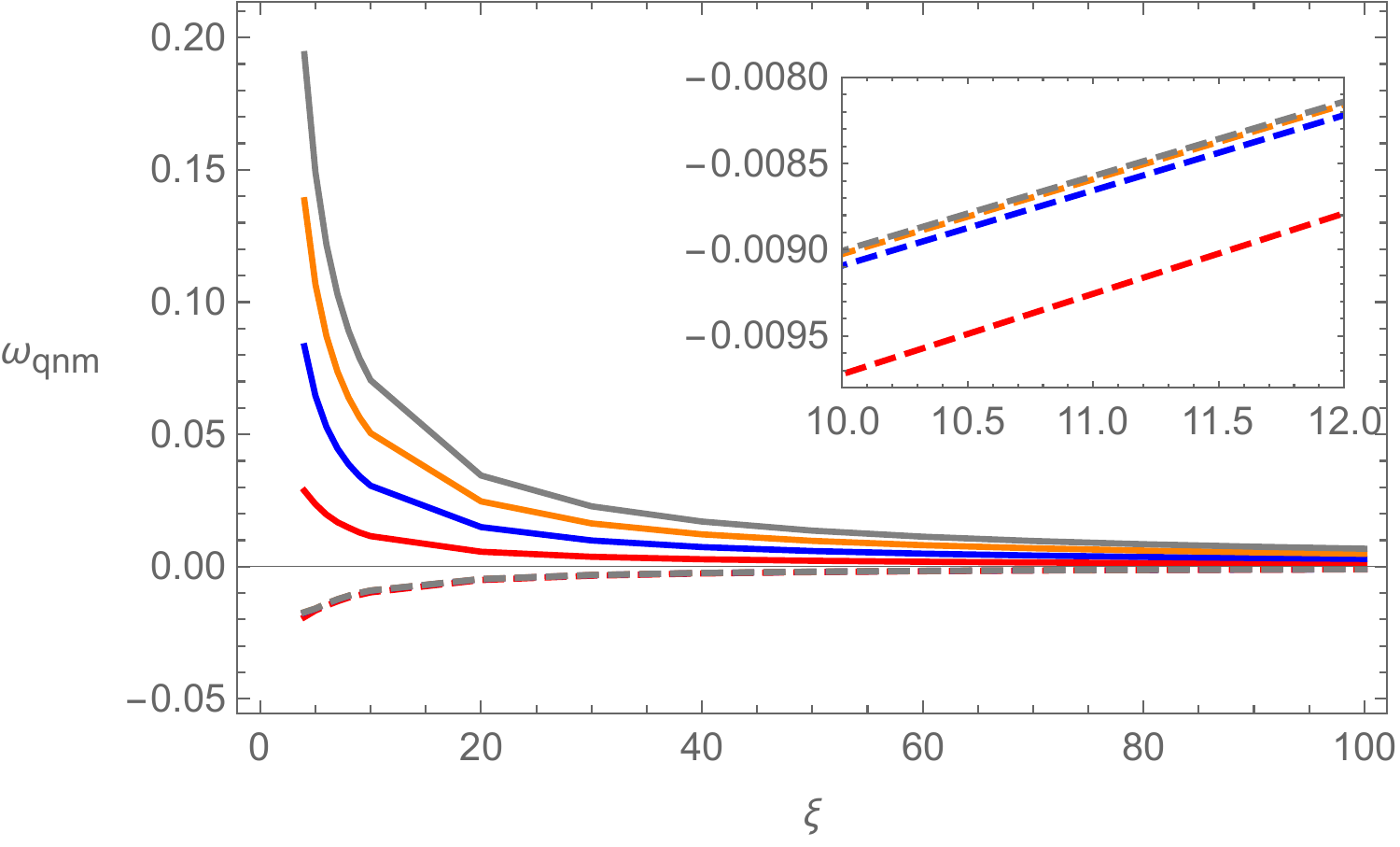}
\caption{The QNM frequency as a function of $\xi$ for different overtone numbers and angular numbers. The solid lines represent $Re(\omega_{qnm})$ and dashed lines represent $Im(\omega_{qnm})$. In the left plot we fix $l=1$. The red, blue, orange and gray lines show the cases $n=0,1,2,4$, respectively; In the right plot, we fix $n=0$. The red, blue, orange and gray lines show the cases $l=0,1,2,3$, respectively.}\label{qnm1}}
\end{figure}

In Fig. \ref{qnm1}, we study the QNMs as the function of the tuning parameter $\xi$ with different overtone numbers $n$ and angular numbers $l$, respectively. The solid lines represent $Re (\omega_{qnm})$ and the dashed lines show $Im (\omega_{qnm})$. It is obvious that all lines approach to zero with the increasing of $\xi$ but never change their sign, which indicates that the acoustic black hole is stable under the perturbations. In detail, in the left plot with fixed $l=1$, the imaginary part is more noticeable, whereas the real part is more slightly different for small $\xi$. It is shown that both $Re(\omega_{qnm})$ and $Im(\omega_{qnm})$ are suppressed for larger $n$, indicating that the perturbation for larger $n$ dies off more quickly. In the right plot with fixed $n=0$,  the real part is obvious for different $l$ and the slight difference of $Im(\omega_{qnm})$ is enlarged in the inset. Moreover, both of $Re(\omega_{qnm})$ and $Im(\omega_{qnm})$ are enhanced with the increasing of $l$. This implies that the perturbation decays faster for smaller angular number $l$. Here the effect of $l$ and $n$ on $\omega_{qnm}$ is qualitatively consistent with  that in RN black hole\cite{Konoplya:2002wt}.

\begin{figure}[thbp]
\center{
\includegraphics[scale=0.5]{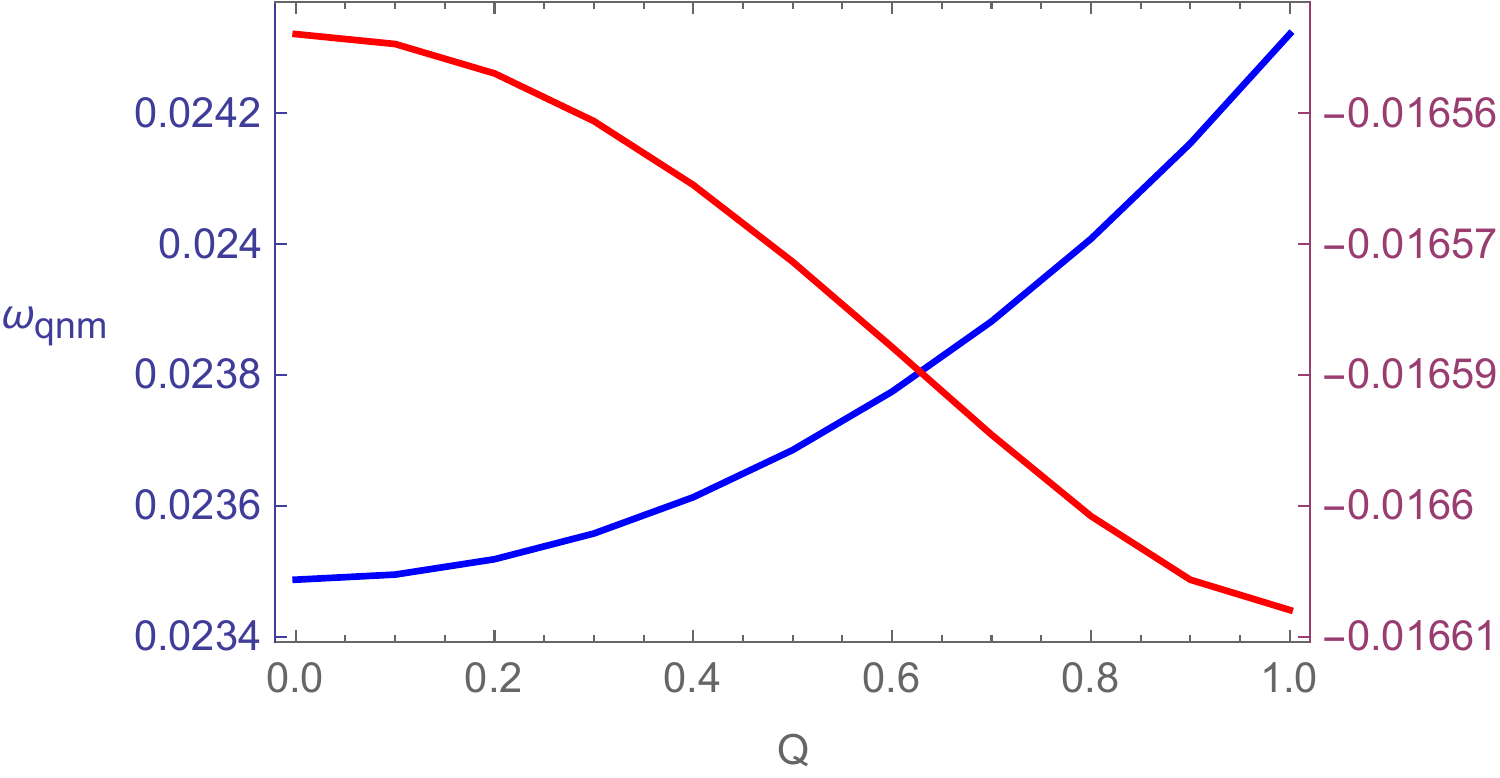}
\caption{The QNM frequency as a function of $Q$ with $\xi=5$ and $n=l=0$. The blue curve represents $Re(\omega_{qnm})$ with the  left Y-axis and the red line represents $Im(\omega_{qnm})$ with the right Y-axis.}\label{qnm2}}
\end{figure}

We also study the effect of $Q$ on the QNM frequency with fixed $\xi=5$ and $n=l=0$. The result is shown in Fig. \ref{qnm2}, in which the blue and red curves represent the real part and imaginary part of the QNM frequency, respectively. The adjustment of $Q$ will not change the sign of $Re(\omega_{qnm})$ or $Im(\omega_{qnm})$, meaning the stability should not be destroyed. But the increasing of $Q$ can  slightly enhance the magnitudes of QNM frequency. This indicates that the oscillation frequency is enhanced and the perturbation dies out quicker. Note that the effect of $Q$ on the QNM frequencies in acoustic black hole is very different from that in RN black hole, in which the QNM frequencies does not change monotonically with $Q$\cite{Konoplya:2002wt,Richartz:2014jla,Cho:2011sf}.

\subsection{QNM frequencies in eikonal limit and acoustic shadow }
In \cite{Cardoso:2008bp}, Cardoso \emph{et al} proposed that the QNM frequencies in the eikonal limit ($l\gg 1$) for a stationary,
spherically symmetric black hole with redshift function $f(r)$  in flat spacetime can be determined by
\begin{eqnarray}
&&\omega_{l \gg 1}=l\ \Omega_{c}-i\ (n+\frac{1}{2})\ |\lambda|,~~~ \mathrm{with}\\
&&\Omega_c=\frac{1}{r_{sh}},~~~~
\lambda=\sqrt{\frac{f(r_{ph})[2 f(r_{ph})-r_{ph}^{2} f^{\prime \prime}(r_{ph})]}{2 r_{ph}^{2}}},\label{eq-wshadow}
\end{eqnarray}
where $n$ is the overtone number. Here the angular velocity $\Omega_c$ and the Lyapunov exponent $\lambda$ are parameters of
the unstable circular null geodesics around the black hole, which can be evaluated via the radius of photon sphere $r_{ph}$ and black hole shadow $r_{sh}$. An extended study could be seen in \cite{Konoplya:2017wot,Jusufi:2019ltj,Cuadros-Melgar:2020kqn,Cai:2020igv,Li:2021zct} and therein.

In this subsection, as a first attempt we shall employ the above proposal to evaluate the QNM frequencies in the eikonal limit ($l\gg 1$) of the acoustic charged black hole which is also stationary, spherically symmetric. This study is motivated from two aspects. On one hand, the numeric results in Fig. \ref{qnm1} show that as $l$ increases, the imaginary part of QNM frequency tends to the horizonal axis. So it is important to check the (in)stability of the acoustic charged black hole in the eikonal limit. On the other hand, analogue to the study of optical black hole, we have studied the acoustic sphere and acoustic shadow of our background in section \ref{sec=shadow}. So we are well prepared, i.e. we could treat in \eqref{eq-wshadow} $r_{ph}$ as the radius of acoustic sphere $r_{ah}$, and $r_{sh}$ as the radius of acoustic shadow for the acoustic charged black hole with $f(r)=\mathcal{F}(r)$ as defined in \eqref{acousticMetrc2}.

\begin{figure}[thbp]
\center{
\includegraphics[scale=0.5]{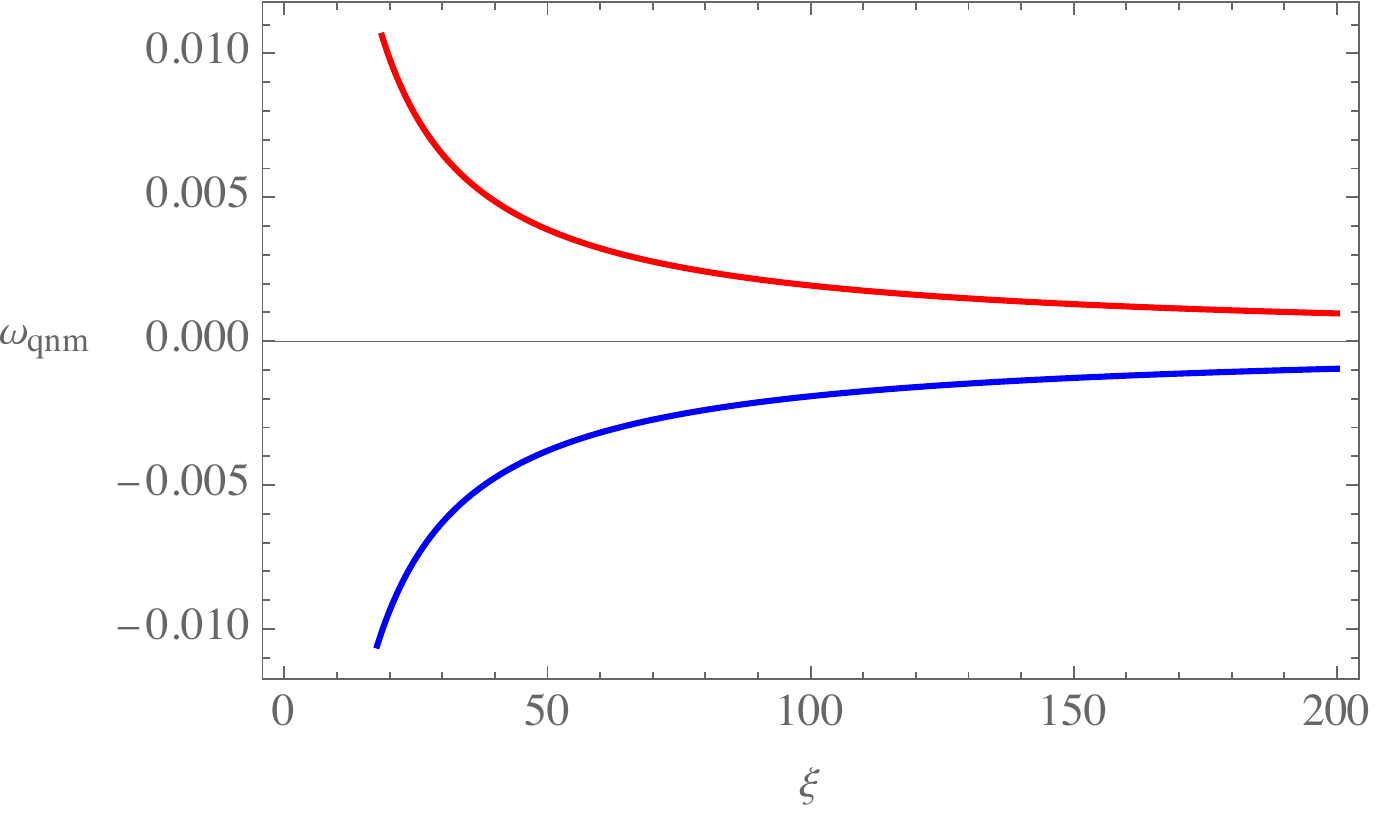}\hspace{1cm}
\caption{The QNM frequency in the eikonal limit as a function of $\xi$ . The red line represents $Re(\omega_{qnm})$ and the blue line represents $Im(\omega_{qnm})$. Here we set $Q=1/2$ and $n=0$.
}\label{qnmeki}}
\end{figure}

Then by substituting $r_{ah}$ and $r_{sh}$ evaluated in section \ref{sec=shadow} into the expression \eqref{eq-wshadow}, we show the QNM frequencies with $n=0$ in the eikonal limit in Fig.\ref{qnmeki}. The dependence of $\omega_{qnm}$ in the eikonal limit on the tuning parameter is consistent with  numerical result in Fig. \ref{qnm1}, namely, that as $\xi$ increases, both $Re(\omega_{qnm})$ and $Im(\omega_{qnm})$ approach to the horizonal axis. More importantly, they could not cross the axis or  change sign. Note that here we set $Q=1/2$, but other valid $Q$ will not affect the qualitative behavior. Thus, the result in eikonal limit further indicates that the acoustic charged black hole keeps stable under the scalar field perturbation.

\section{Analogue Hawking radiation}\label{sec=GBfactor}
To further study the near-horizon features, we shall study the analogue Hawking radiation for our acoustic charged black hole. Similar to the case of astrophysical black hole, there is the analogue Hawking radiation which emits thermal flux of particles, and the gradient of the velocity field at the acoustic horizon gives the acoustic temperature.
Basically, the Hawking radiation process is related to a scattering problem, because the particles emitted from the black hole in the vicinity of the horizon cannot penetrate the potential barrier if they do not have enough energy and only part of them can be observed far away from the near-horizon area. Note that the effective potential barrier decays monotonically both at acoustic horizon and infinity just shown in Fig. \ref{potential}, thus, it is convenient to use the WKB approach to study the grey-body factor which would give us the transmission of particles through the effective potential and the energy radiation rate of the acoustic black hole.

To proceed, we shall solve the wave equation Eq.(\ref{eq1}) by considering the scattering boundary condition which allows incoming waves from infinity,
\begin{align}
&\Psi=T e^{-i\omega r_\ast}, \quad \quad \quad r_\ast \rightarrow -\infty,\\
&\Psi=e^{-i\omega r_\ast}+Re^{i\omega r_\ast}, \quad \quad \quad r_\ast \rightarrow +\infty.
\end{align}
Here $R$ and $T$ are the reflection and transmission coefficient, respectively, which satisfy $|T|^2+|R|^2=1$. For each angular number $l$, the grey-body factor can be given by the transmission coefficient as\cite{Iyer:1986np}
\begin{equation}
|\mathcal{A}_l|^2=1-|R_l|^2=|T_l|^2\quad \mathrm{and}\quad |T_l|^2=(1+e^{2i\pi K})^{-1},
\end{equation}
where $K$ is determined by
\begin{equation}\label{eq-K}
K=i\frac{\omega^2-V_0}{\sqrt{-2V_0''}}-\sum_{i=2}^{i=6}\Lambda_i(K).
\end{equation}
In the above equation, $V_0$ denotes the maximum value of the effective potential and and $V_0''$ represents its second-order derivative with respect to the tortoise coordinates; $\Lambda_i$ are the higher WKB corrections which depend on the maximum of the 2$i$th order derivatives of $K$ and the potential\cite{Schutz:1985km,Iyer:1986np,Konoplya:2003ii}.

As addressed in  \cite{Hawking:1974sw}, one can obtain the energy emission rate in term of the grey-body factor via
\begin{equation}\label{ene1}
\frac{dE}{dt}=\sum_{l}N_l|\mathcal{A}_l|^2\frac{\omega}{e^{\omega/T_H}-1}\frac{d\omega}{2\pi},
\end{equation}
where the Hawking temperature is defined as $T_H=-\mathcal{F}'(r_{ac})/4\pi$ , and for the scalar field, the multiplicities satisfy $N_l=2l+1$.
Then, we employ the 6th order WKB method to calculate the grey-body factor and energy emission rate of the analogue Hawking radiation as the function of frequency. Then we shall study the effects of different parameters on the analogue Hawking radiations.

\begin{figure}[thbp]
\center{
\includegraphics[scale=0.5]{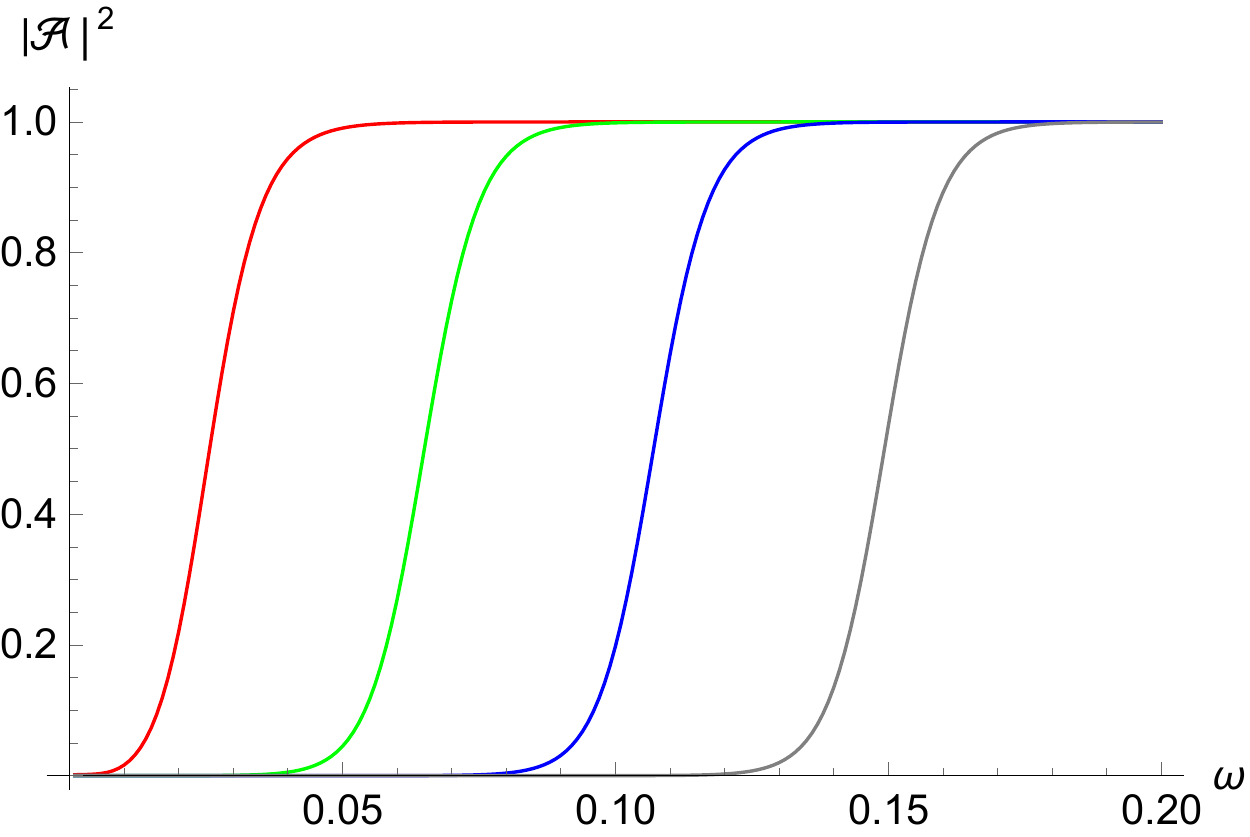}\hspace{1cm}
\includegraphics[scale=0.5]{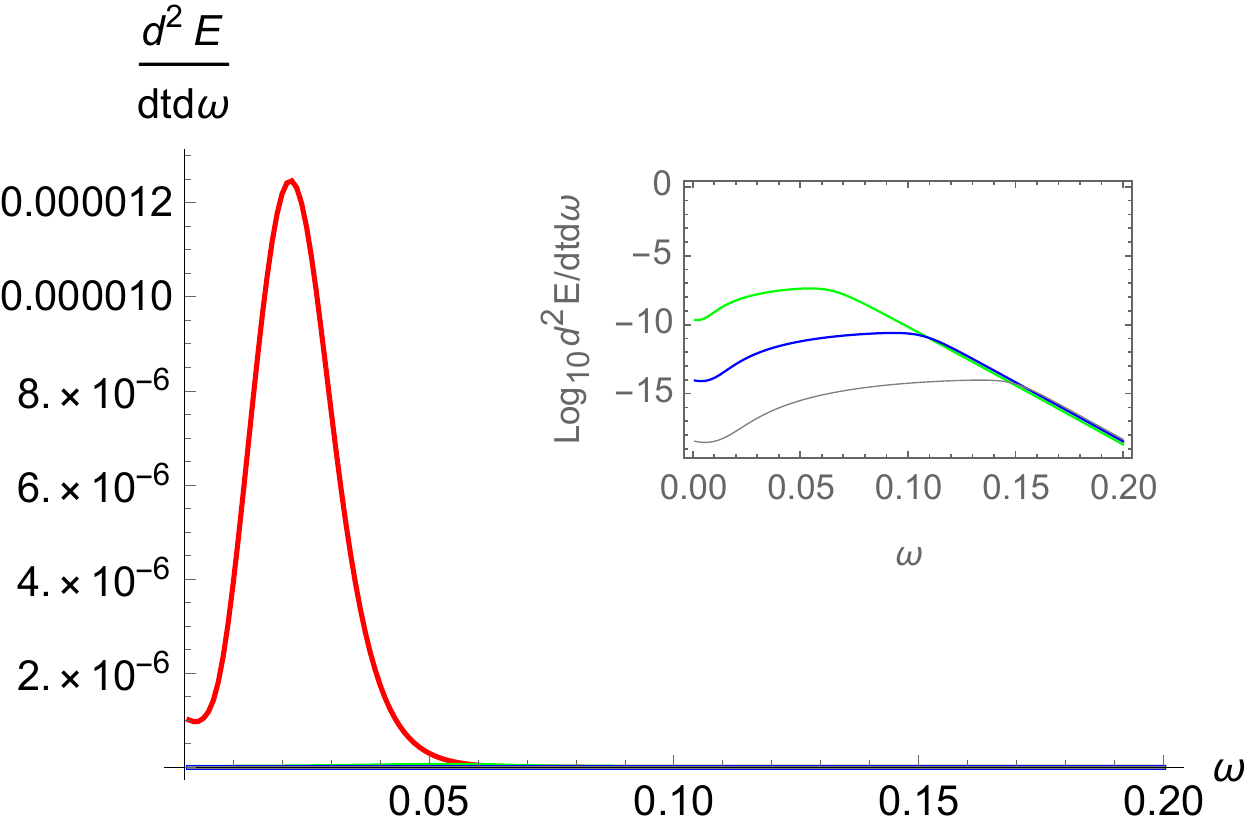}
\caption{The left panel shows the grey-body factor  and the right panel shows the partial energy radiation rate for different angular numbers. For both panels, we fix $\xi=5$ and $Q=1/2$, and the red, green, blue and gray lines correspond to angular numbers $l=0,1, 2$ and  $3$, respectively.  On the right plot, the radiation rate for $l>0$ are too weak to be observable, {so we insert the log plot to distinguish the profiles.}}\label{grfactor1}}
\end{figure}

In Fig. \ref{grfactor1}, with fixed $\xi=5$ and $Q=1/2$, we show the grey-body factor and related energy emission with different angular numbers $l$.  In the left plot, higher frequency represents larger energy of the particles, they are more likely to penetrate the potential barrier and this fact explains why the grey-body factor monotonically increases with the frequency. Moreover, lower angular number would lead to faster saturation of the grey body factor as we show. This behavior is easy to understand as the effective potential barrier grows as $l$ increases shown in right panel of  Fig. \ref{potential}, which means that particles need more energy to penetrate the potential barrier. In the right plot, we find that the mode $l=0$ dominates the analogue Hawking radiation, and the contribution of higher order of $l$ is very small{ whose log plot is inserted. It is obvious that the contributions of higher order of $l$ decay exponentially and could be negligible.} This behavior is consistent with the results in the acoustic Schwarzschild black hole case \cite{Guo:2020blq}.

\begin{figure}[thbp]
\center{
\includegraphics[scale=0.5]{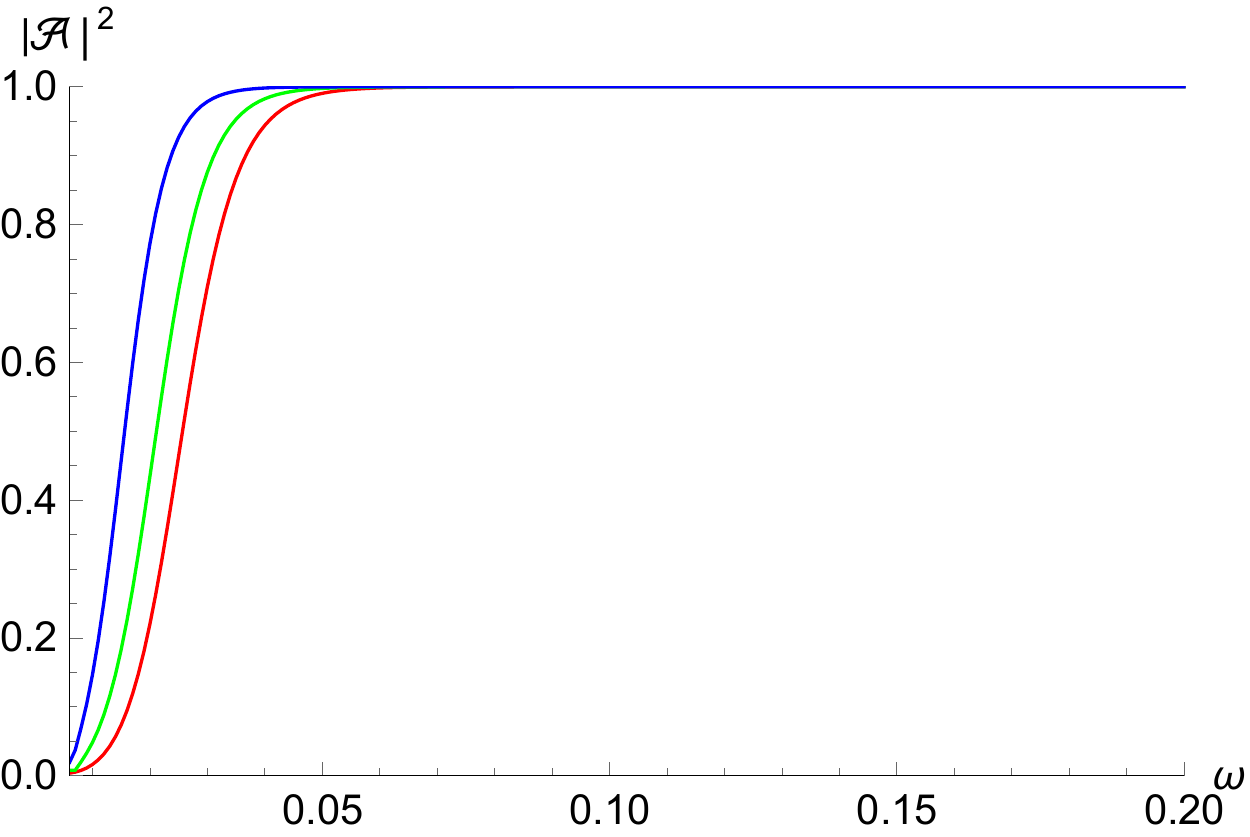}\hspace{1cm}
\includegraphics[scale=0.5]{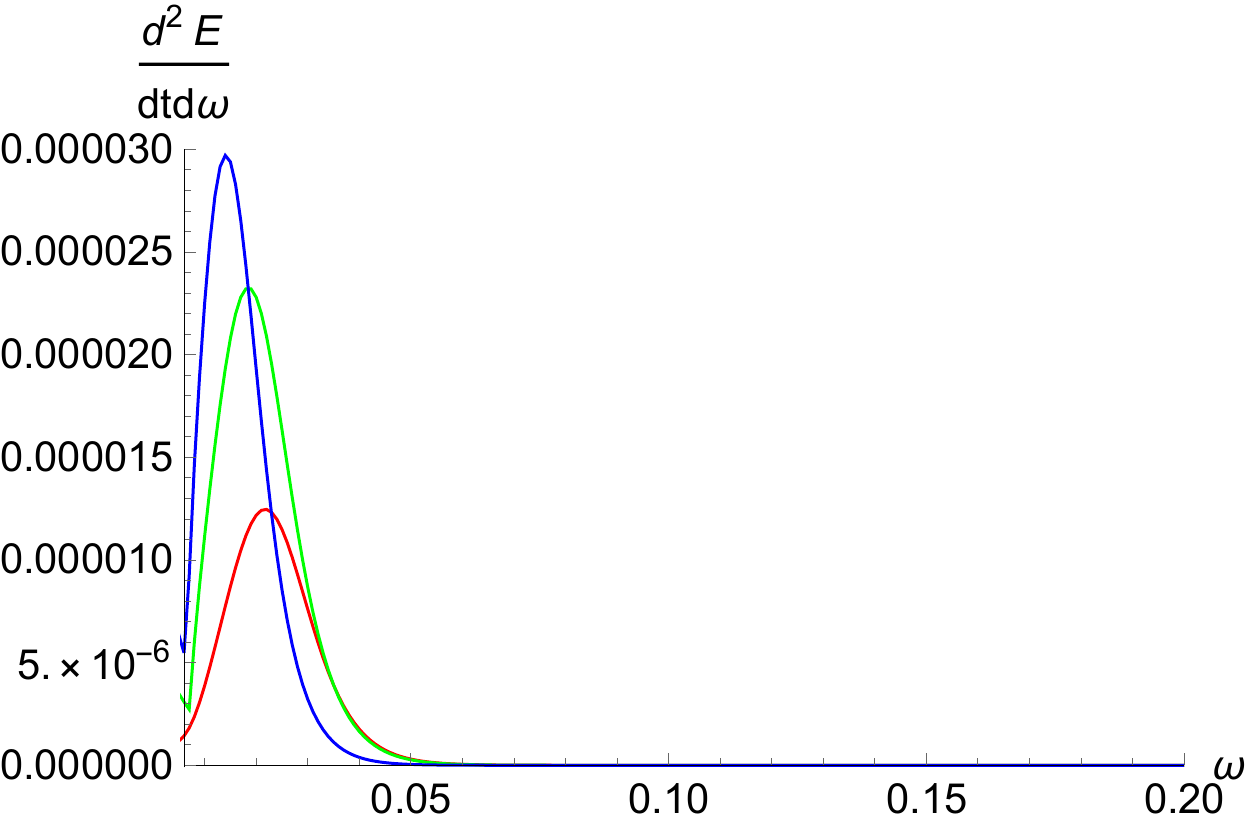}
\caption{The left panel shows the grey-body factor and the right panel shows the partial energy radiation rate  for different $\xi$. For both panels, we fix $l=0$ and $Q=1/2$, and the red line, green line and blue line correspond to  $\xi=5$, $\xi=6$ and $\xi=8$, respectively.}\label{grfactor2}}
\end{figure}

\begin{figure}[thbp]
\center{
\includegraphics[scale=0.5]{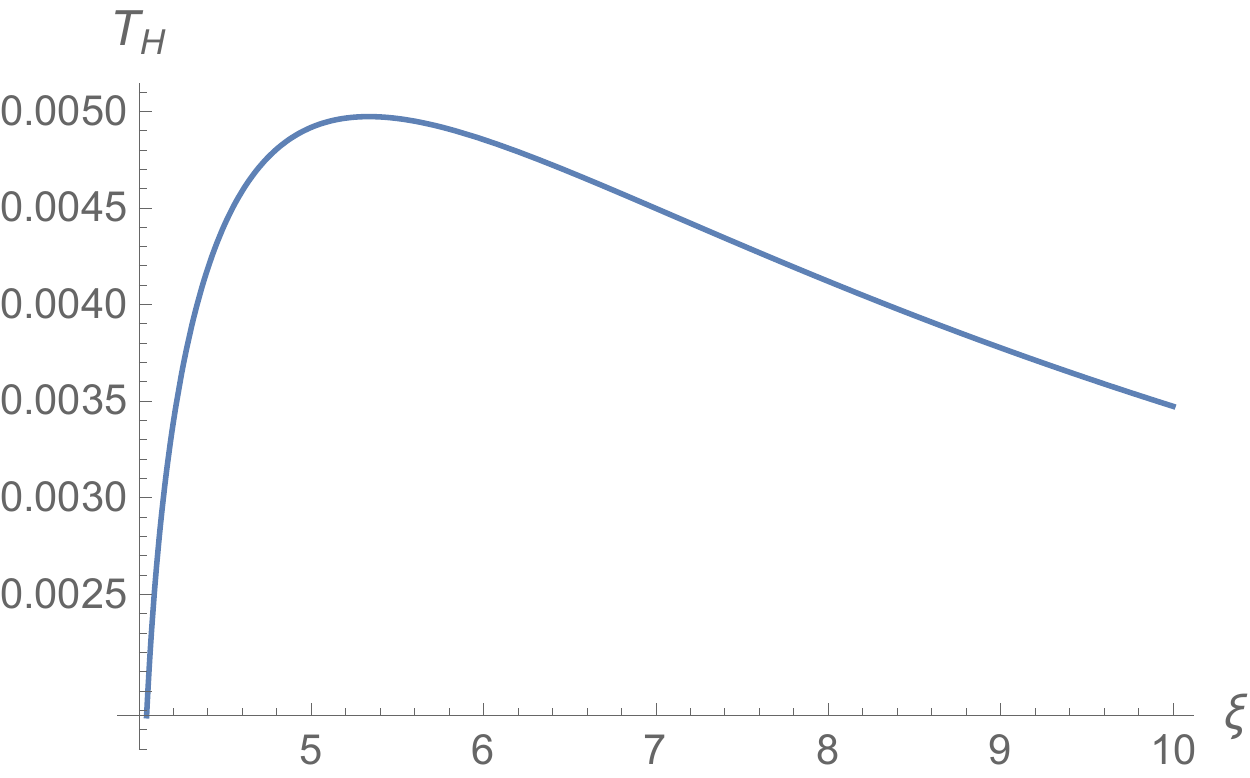}\hspace{1cm}
\includegraphics[scale=0.5]{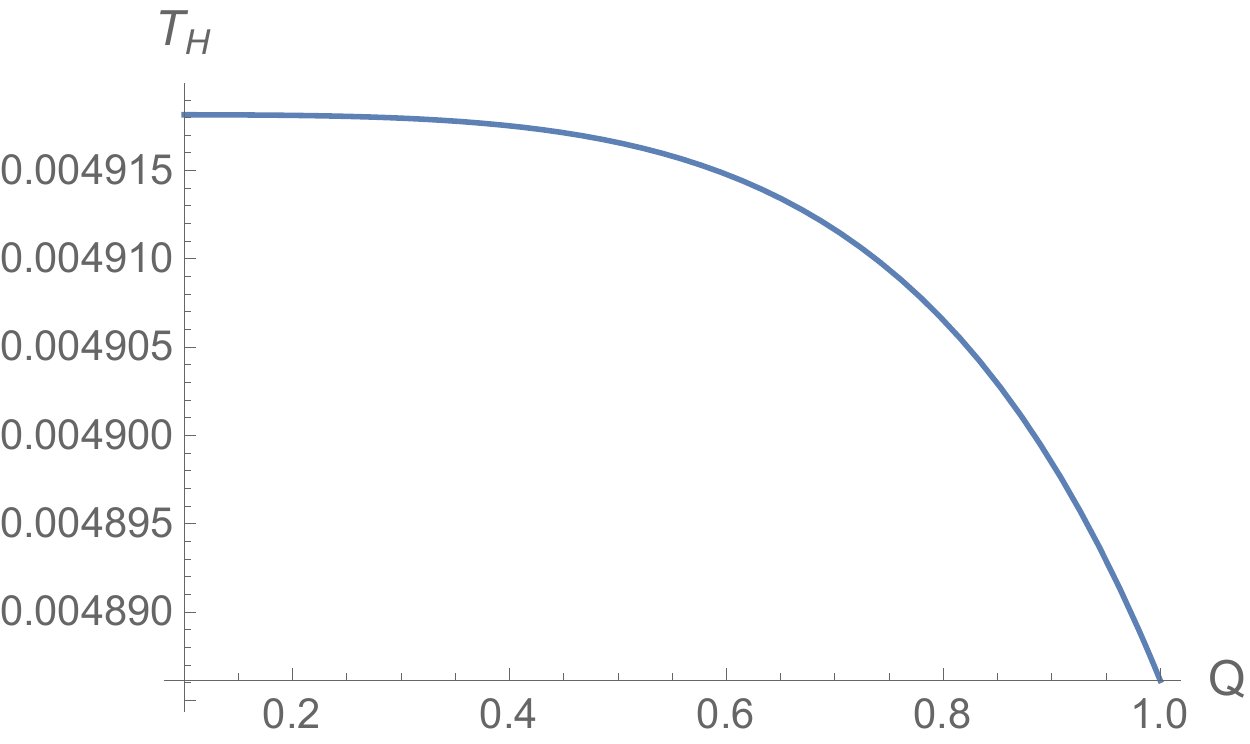}
\caption{The behavior of Hawking temperature as a function of $\xi$ and $Q$. Here we set $Q=1/2$ in the left while $\xi=5$ in the right.}\label{temp}}
\end{figure}

In Fig. \ref{grfactor2}, by fixing $l=0$ and $Q=1/2$, we show the effect of the tuning parameter $\xi$. The results are consistent with the analysis of the effective potential shown in middle panel of Fig. \ref{potential}. In the left plot, the grey-body factor is enhanced when the tuning parameter $\xi$ increases, this is because the potential barrier decreases with the increasing of $\xi$. In the right plot, the energy emission rate at the low frequency region is also enhanced by  increasing $\xi$. But in the higher frequency region, this behavior will change because of the dependence of the Hawking  temperature. From the left panel of Fig. \ref{temp}, one can observe that the Hawking temperature goes from zero at $\xi=4$ and reaches the maximum at $\xi=5.334$, and then the falloff  of the Hawking temperature causes the suppression of the energy emission rate.
\begin{figure}[thbp]
\center{
\includegraphics[scale=0.5]{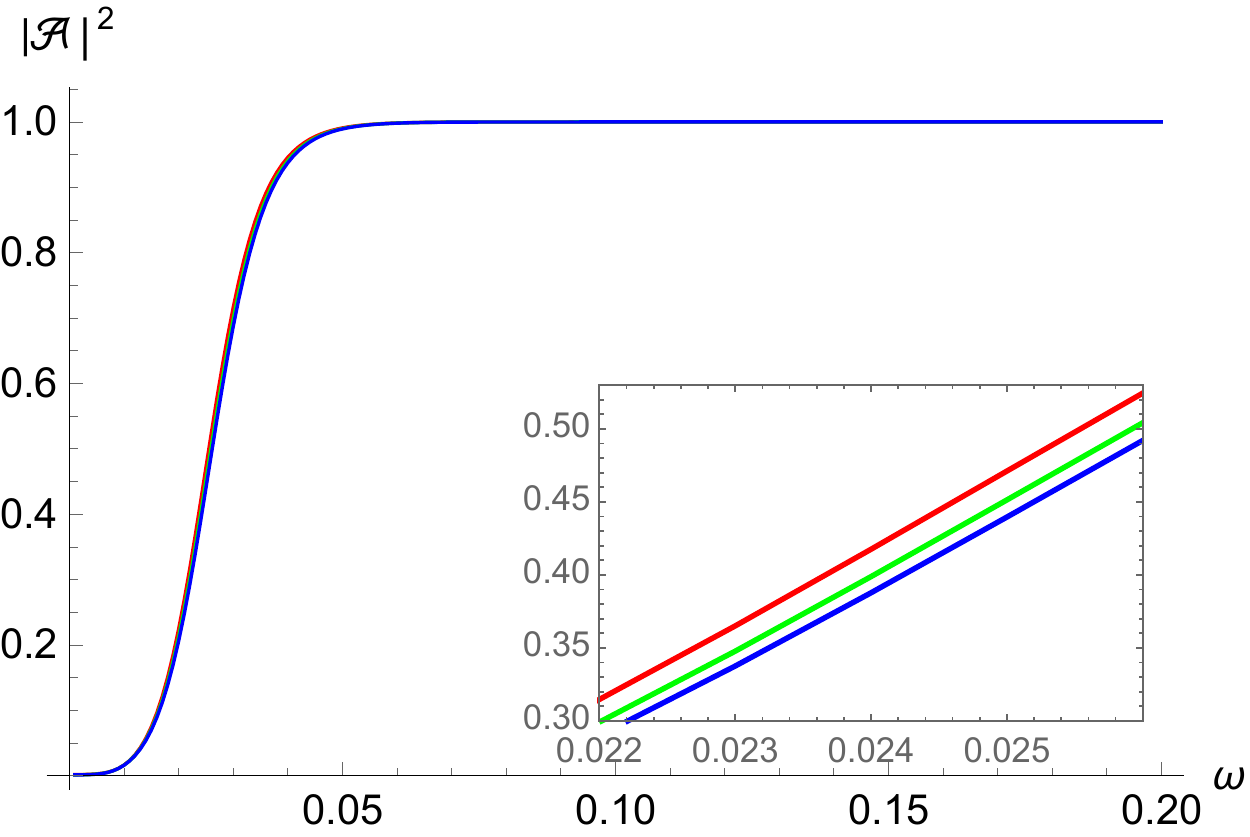}\hspace{1cm}
\includegraphics[scale=0.5]{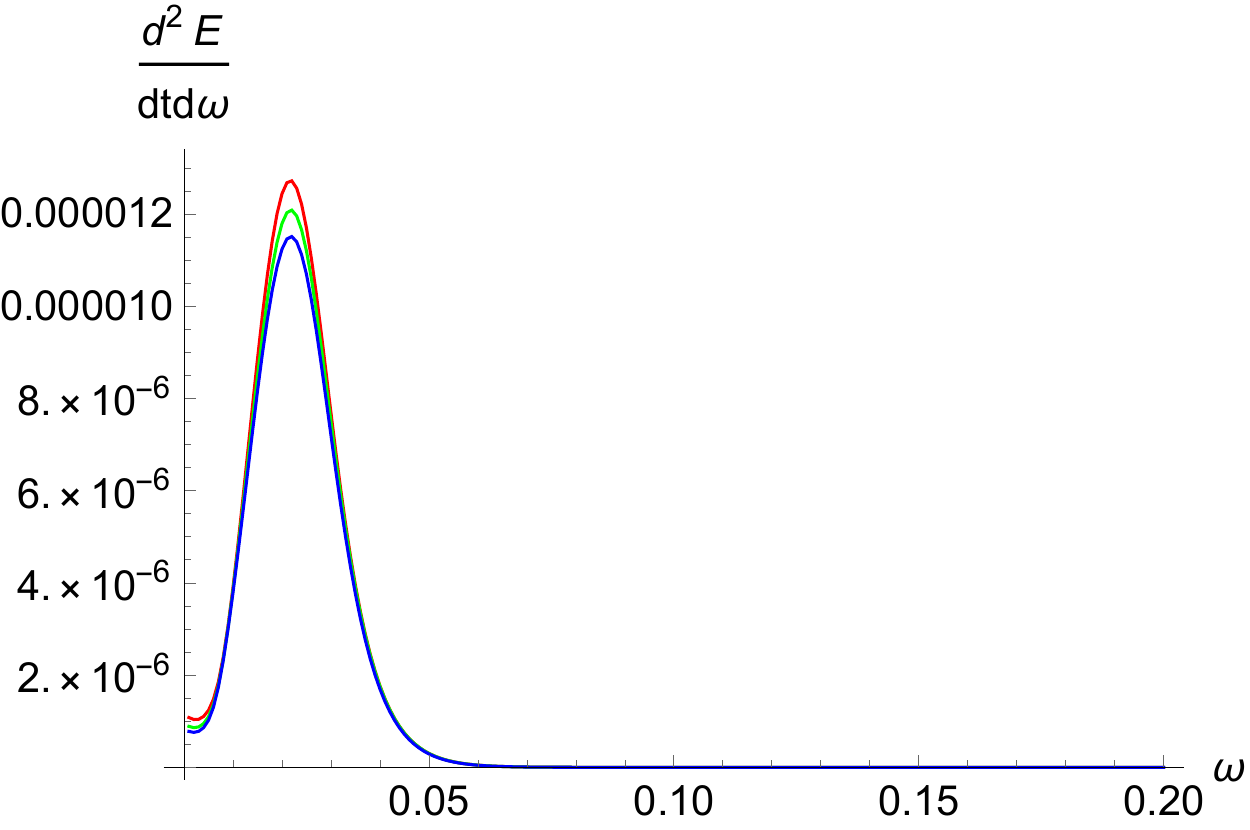}
\caption{The left panel shows the grey-body factor and the right panel shows the partial energy radiation rate  for different $Q$. For both panels, we fix $l=0$ and $\xi=5$, and the red line, green line and blue line correspond to  $Q=0.1$, $Q=0.8$ and $Q=1$, respectively.}\label{grfactor3}}
\end{figure}

We then move on to study the effect of  black hole charge $Q$ on the grey-body factor and related energy emission rate by fixing $l=0$ and $\xi=5$. The left panel of Fig. \ref{grfactor3} shows the behavior of the grey-body factor as the function of frequency $\omega$, which is suppressed slightly by the larger $Q$. The energy emission rate is consistent with the grey-body factor, its peak decreases with the increase of $Q$ in the low frequency region. Moreover, we also discuss the Hawking temperature as the function of the Q when it is shown that the extreme black hole with $Q=1$ has the  minimal analogue Hawking temperature in the right of Fig. \ref{temp}. The monotonic behavior of the Hawking temperature supports the behavior of the energy emission rate.

\section{Conclusion and discussion}
In this paper, we constructed the `curved' acoustic black hole with charges, and then explored the near-horizon characteristics after analyzing its horizon structures.
Firstly, we studied the acoustic black hole shadow as analogue with the optical shadow in astrophysical black holes. Our results show that the radius of the acoustic shadow is enhanced with the increasing of the tuning parameter $\xi$ but is suppressed by the black hole charge $Q$. The increase of acoustic shadow  almost linearly increases with tuning parameter $\xi$. This is reasonable because the acoustic horizon is larger for larger $\xi$.

We then investigated the QNM frequency and the analogue Hawking radiation by considering a massless scalar field as the perturbation of the acoustic charged background. It was shown that the acoustic charged black hole is stable under the scalar field perturbation as the imaginary part of QNM frequency is always negative. As we increase the tuning parameter $\xi$, both the positive real part and negative imaginary part of QNM frequency approach to zero but never change signs, which suggests that the oscillation of the perturbation is suppressed while the damping time of the perturbation is longer for larger $\xi$.
The signal of the acoustic black hole QNM frequency are much weaker than the astrophysical black hole. This indicates that it is more likely to observe the signal of the acoustic horizon in the universe.
We also studied the effect of the black hole charge $Q$. As $Q$ increases, the oscillation is enhanced slightly and the perturbation decays a little faster. This behavior is very different from that in RN black hole as disclosed in \cite{Konoplya:2002wt,Richartz:2014jla,Cho:2011sf} that the QNM frequencies presented a non-monotonic behavior as  $Q$.

Moreover, we also computed the QNM frequencies in the eikonal limit in terms of the acoustic shadow by following  Cardoso's  \emph{et al} proposal in \cite{Cardoso:2008bp}. It was found that in the eikonal limit,  the QNM frequencies approach to zero but never change their sign as $\xi$ increases. This phenomenon holds for different black hole charges $Q$, meaning further that the acoustic charged black hole is stable under the scalar perturbation. It is noted that to further study the near-horizon properties and test the (in)stability, other perturbations such as gravitational perturbation or electromagnetic perturbation have to be considered, which we shall present elsewhere.

Finally, we studied the analogue Hawking radiation of the acoustic charged black hole. Both of the grey-body factor and the energy emission rate are suppressed by larger angular number $l$ or black hole charge $Q$ which is attributed to the growing of the effective potential barrier. Meanwhile, as the tuning parameter $\xi$ increases, the grey-body factor is enhanced, but the energy emission rate is not monotonic depending on the frequencies, which results from the
nonmonotonicity of the analogue Hawking temperature on $\xi$.

It is noticed that in last decades, significant progress about the analogue
black holes emerged from Minkowski spacetime have been made in both theoretical and experimental sides. Further experimental simulation of the acoustic black hole could help us to understand the astrophysical phenomena and the near-horizon properties of the real black holes. Theoretically, acoustic black hole in curved spacetime is attracting widespread attention and interest as it is more realistic, especially some analogue research has been done in acoustic Schwarzschild black hole \cite{Ge:2019our,Guo:2020blq,Vieira:2021xqw}. Along this line, besides the current study, more interesting outcomes in curved acoustic black holes deserve further investigations, for instance, \emph{i)} Further consider the electromagnetic and analogue gravitational perturbations and  test their (in)stability; \emph{ ii)} It is natural to extend the curved acoustic black hole into a rotating case and study the properties of curved acoustic rotating black hole; \emph{iii)} Another attempt is to discuss the curved acoustic black hole in the wormhole background; \emph{iv)} It was addressed in \cite{Sun:2017eph} that the acoustic black hole can also be realized by the holographic approach. This gives more interest to relate the curved acoustic black hole with the holographic principle \cite{Hartnoll:2008vx} and holographic vortex \cite{Domenech:2010nf,Montull:2011im,Salvio:2012at,Bao:2013fda,Guo:2014wca}; \emph{ v)} Next but not the last interesting issue is how to mimic the charge in the experiment since in the current theoretical setup, the fluid is not charged and the 'charge' is introduced in its velocity.

\begin{acknowledgments}
We appreciate Yen Chin Ong for helpful discussion. This work is supported by the Natural Science Foundation
of China under Grant Nos.11705161, 12075202 and 11690022. Xiao-Mei Kuang is also supported by Fok Ying Tung Education Foundation under grant No.171006 and Natural
Science Foundation of Jiangsu Province under Grant No.BK20211601.
\end{acknowledgments}

\bibliographystyle{JHEP}
\bibliography{AcusBH}

\providecommand{\href}[2]{#2}\begingroup\raggedright\begin{thebibliography}{10}

\bibitem{TheLIGOScientific:2016wfe}
{\scshape LIGO Scientific, Virgo} collaboration, B.~P. Abbott et~al.,
  \emph{{Properties of the Binary Black Hole Merger GW150914}},
  \href{http://dx.doi.org/10.1103/PhysRevLett.116.241102}{\emph{Phys. Rev.
  Lett.} {\bf 116} (2016) 241102}, [\href{http://arxiv.org/abs/1602.03840}{{\tt
  1602.03840}}].

\bibitem{Akiyama:2019cqa}
{\scshape Event Horizon Telescope} collaboration, K.~Akiyama et~al.,
  \emph{{First M87 Event Horizon Telescope Results. I. The Shadow of the
  Supermassive Black Hole}},
  \href{http://dx.doi.org/10.3847/2041-8213/ab0ec7}{\emph{Astrophys. J. Lett.}
  {\bf 875} (2019) L1}, [\href{http://arxiv.org/abs/1906.11238}{{\tt
  1906.11238}}].

\bibitem{Akiyama:2019brx}
{\scshape Event Horizon Telescope} collaboration, K.~Akiyama et~al.,
  \emph{{First M87 Event Horizon Telescope Results. II. Array and
  Instrumentation}},
  \href{http://dx.doi.org/10.3847/2041-8213/ab0c96}{\emph{Astrophys. J. Lett.}
  {\bf 875} (2019) L2}, [\href{http://arxiv.org/abs/1906.11239}{{\tt
  1906.11239}}].

\bibitem{Akiyama:2019sww}
{\scshape Event Horizon Telescope} collaboration, K.~Akiyama et~al.,
  \emph{{First M87 Event Horizon Telescope Results. III. Data Processing and
  Calibration}},
  \href{http://dx.doi.org/10.3847/2041-8213/ab0c57}{\emph{Astrophys. J. Lett.}
  {\bf 875} (2019) L3}, [\href{http://arxiv.org/abs/1906.11240}{{\tt
  1906.11240}}].

\bibitem{Akiyama:2019bqs}
{\scshape Event Horizon Telescope} collaboration, K.~Akiyama et~al.,
  \emph{{First M87 Event Horizon Telescope Results. IV. Imaging the Central
  Supermassive Black Hole}},
  \href{http://dx.doi.org/10.3847/2041-8213/ab0e85}{\emph{Astrophys. J. Lett.}
  {\bf 875} (2019) L4}, [\href{http://arxiv.org/abs/1906.11241}{{\tt
  1906.11241}}].

\bibitem{Akiyama:2019fyp}
{\scshape Event Horizon Telescope} collaboration, K.~Akiyama et~al.,
  \emph{{First M87 Event Horizon Telescope Results. V. Physical Origin of the
  Asymmetric Ring}},
  \href{http://dx.doi.org/10.3847/2041-8213/ab0f43}{\emph{Astrophys. J. Lett.}
  {\bf 875} (2019) L5}, [\href{http://arxiv.org/abs/1906.11242}{{\tt
  1906.11242}}].

\bibitem{Akiyama:2019eap}
{\scshape Event Horizon Telescope} collaboration, K.~Akiyama et~al.,
  \emph{{First M87 Event Horizon Telescope Results. VI. The Shadow and Mass of
  the Central Black Hole}},
  \href{http://dx.doi.org/10.3847/2041-8213/ab1141}{\emph{Astrophys. J. Lett.}
  {\bf 875} (2019) L6}, [\href{http://arxiv.org/abs/1906.11243}{{\tt
  1906.11243}}].

\bibitem{Press:1971wr}
W.~H. Press, \emph{{Long Wave Trains of Gravitational Waves from a Vibrating
  Black Hole}}, \href{http://dx.doi.org/10.1086/180849}{\emph{Astrophys. J.
  Lett.} {\bf 170} (1971) L105--L108}.

\bibitem{Vishveshwara:1970zz}
C.~V. Vishveshwara, \emph{{Scattering of Gravitational Radiation by a
  Schwarzschild Black-hole}},
  \href{http://dx.doi.org/10.1038/227936a0}{\emph{Nature} {\bf 227} (1970)
  936--938}.

\bibitem{Hawking:1974rv}
S.~W. Hawking, \emph{{Black hole explosions}},
  \href{http://dx.doi.org/10.1038/248030a0}{\emph{Nature} {\bf 248} (1974)
  30--31}.

\bibitem{Hawking:1974sw}
S.~Hawking, \emph{{Particle Creation by Black Holes}},
  \href{http://dx.doi.org/10.1007/BF02345020}{\emph{Commun. Math. Phys.} {\bf
  43} (1975) 199--220}.

\bibitem{Unruh:1980cg}
W.~G. Unruh, \emph{{Experimental black hole evaporation}},
  \href{http://dx.doi.org/10.1103/PhysRevLett.46.1351}{\emph{Phys. Rev. Lett.}
  {\bf 46} (1981) 1351--1353}.

\bibitem{Visser:1997ux}
M.~Visser, \emph{{Acoustic black holes: Horizons, ergospheres, and Hawking
  radiation}}, \href{http://dx.doi.org/10.1088/0264-9381/15/6/024}{\emph{Class.
  Quant. Grav.} {\bf 15} (1998) 1767--1791},
  [\href{http://arxiv.org/abs/gr-qc/9712010}{{\tt gr-qc/9712010}}].

\bibitem{Cardoso:2004fi}
V.~Cardoso, J.~P.~S. Lemos and S.~Yoshida, \emph{{Quasinormal modes and
  stability of the rotating acoustic black hole: Numerical analysis}},
  \href{http://dx.doi.org/10.1103/PhysRevD.70.124032}{\emph{Phys. Rev. D} {\bf
  70} (2004) 124032}, [\href{http://arxiv.org/abs/gr-qc/0410107}{{\tt
  gr-qc/0410107}}].

\bibitem{Chen:2004zr}
S.-b. Chen and J.-l. Jing, \emph{{Asymptotic quasinormal modes of a coupled
  scalar field in the Garfinkle-Horowitz-Strominger dilaton spacetime}},
  \href{http://dx.doi.org/10.1088/0264-9381/22/3/006}{\emph{Class. Quant.
  Grav.} {\bf 22} (2005) 533--540},
  [\href{http://arxiv.org/abs/gr-qc/0409013}{{\tt gr-qc/0409013}}].

\bibitem{Benone:2014nla}
C.~L. Benone, L.~C.~B. Crispino, C.~Herdeiro and E.~Radu, \emph{{Acoustic
  clouds: standing sound waves around a black hole analogue}},
  \href{http://dx.doi.org/10.1103/PhysRevD.91.104038}{\emph{Phys. Rev. D} {\bf
  91} (2015) 104038}, [\href{http://arxiv.org/abs/1412.7278}{{\tt 1412.7278}}].

\bibitem{Sarkar:2017puh}
S.~Sarkar and A.~Bhattacharyay, \emph{{Quantum Potential induced UV-IR coupling
  in Analogue Hawking radiation: From Bose-Einstein Condensates to canonical
  acoustic black holes}},
  \href{http://dx.doi.org/10.1103/PhysRevD.96.064027}{\emph{Phys. Rev. D} {\bf
  96} (2017) 064027}, [\href{http://arxiv.org/abs/1703.08027}{{\tt
  1703.08027}}].

\bibitem{Liberati:2020mdr}
S.~Liberati, G.~Tricella and A.~Trombettoni, \emph{{Back-reaction in canonical
  analogue black holes}},
  \href{http://dx.doi.org/10.3390/app10248868}{\emph{Appl. Sciences} {\bf 10}
  (2020) 8868}, [\href{http://arxiv.org/abs/2010.09966}{{\tt 2010.09966}}].

\bibitem{Vieira:2014rva}
H.~S. Vieira and V.~B. Bezerra, \emph{{Acoustic black holes: massless scalar
  field analytic solutions and analogue Hawking radiation}},
  \href{http://dx.doi.org/10.1007/s10714-016-2082-x}{\emph{Gen. Rel. Grav.}
  {\bf 48} (2016) 88}, [\href{http://arxiv.org/abs/1406.6884}{{\tt
  1406.6884}}].

\bibitem{Nakano:2004ha}
H.~Nakano, Y.~Kurita, K.~Ogawa and C.-M. Yoo, \emph{{Quasinormal ringing for
  acoustic black holes at low temperature}},
  \href{http://dx.doi.org/10.1103/PhysRevD.71.084006}{\emph{Phys. Rev. D} {\bf
  71} (2005) 084006}, [\href{http://arxiv.org/abs/gr-qc/0411041}{{\tt
  gr-qc/0411041}}].

\bibitem{Barcelo:2005fc}
C.~Barcelo, S.~Liberati and M.~Visser, \emph{{Analogue gravity}},
  \href{http://dx.doi.org/10.12942/lrr-2005-12}{\emph{Living Rev. Rel.} {\bf 8}
  (2005) 12}, [\href{http://arxiv.org/abs/gr-qc/0505065}{{\tt gr-qc/0505065}}].

\bibitem{Berti:2004ju}
E.~Berti, V.~Cardoso and J.~P.~S. Lemos, \emph{{Quasinormal modes and classical
  wave propagation in analogue black holes}},
  \href{http://dx.doi.org/10.1103/PhysRevD.70.124006}{\emph{Phys. Rev. D} {\bf
  70} (2004) 124006}, [\href{http://arxiv.org/abs/gr-qc/0408099}{{\tt
  gr-qc/0408099}}].

\bibitem{Chen:2006zy}
S.-B. Chen and J.-L. Jing, \emph{{Quasinormal modes of a coupled scalar field
  in the acoustic black hole spacetime}},
  \href{http://dx.doi.org/10.1088/0256-307X/23/1/007}{\emph{Chin. Phys. Lett.}
  {\bf 23} (2006) 21--24}.

\bibitem{Anacleto:2019rfn}
M.~A. Anacleto, F.~A. Brito, C.~V. Garcia, G.~C. Luna and E.~Passos,
  \emph{{Quantum-corrected rotating acoustic black holes in Lorentz-violating
  background}},
  \href{http://dx.doi.org/10.1103/PhysRevD.100.105005}{\emph{Phys. Rev. D} {\bf
  100} (2019) 105005}, [\href{http://arxiv.org/abs/1904.04229}{{\tt
  1904.04229}}].

\bibitem{Balbinot:2019mei}
R.~Balbinot, A.~Fabbri, R.~A. Dudley and P.~R. Anderson, \emph{{Particle
  Production in the Interiors of Acoustic Black Holes}},
  \href{http://dx.doi.org/10.1103/PhysRevD.100.105021}{\emph{Phys. Rev. D} {\bf
  100} (2019) 105021}, [\href{http://arxiv.org/abs/1910.04532}{{\tt
  1910.04532}}].

\bibitem{Eskin:2019tin}
G.~Eskin, \emph{{New examples of Hawking radiation from acoustic black holes}},
   \href{http://arxiv.org/abs/1906.06038}{{\tt 1906.06038}}.

\bibitem{Eskin:2019mqi}
G.~Eskin, \emph{{Hawking type radiation from acoustic black holes with
  time-dependent metric}},  \href{http://arxiv.org/abs/1912.12775}{{\tt
  1912.12775}}.

\bibitem{Zhang:2016pqx}
B.~Zhang, \emph{{Thermodynamics of acoustic black holes in two dimensions}},
  \href{http://dx.doi.org/10.1155/2016/5710625}{\emph{Adv. High Energy Phys.}
  {\bf 2016} (2016) 5710625}, [\href{http://arxiv.org/abs/1606.00693}{{\tt
  1606.00693}}].

\bibitem{Wang:2019zqw}
Q.-B. Wang and X.-H. Ge, \emph{{Geometry outside of acoustic black holes in
  ($2+1$)-dimensional spacetime}},
  \href{http://dx.doi.org/10.1103/PhysRevD.102.104009}{\emph{Phys. Rev. D} {\bf
  102} (2020) 104009}, [\href{http://arxiv.org/abs/1912.05285}{{\tt
  1912.05285}}].

\bibitem{Lahav:2009wx}
O.~Lahav, A.~Itah, A.~Blumkin, C.~Gordon and J.~Steinhauer, \emph{{Realization
  of a sonic black hole analogue in a Bose-Einstein condensate}},
  \href{http://dx.doi.org/10.1103/PhysRevLett.105.240401}{\emph{Phys. Rev.
  Lett.} {\bf 105} (2010) 240401}, [\href{http://arxiv.org/abs/0906.1337}{{\tt
  0906.1337}}].

\bibitem{deNova:2018rld}
J.~R. Mu\~noz~de Nova, K.~Golubkov, V.~I. Kolobov and J.~Steinhauer,
  \emph{{Observation of thermal Hawking radiation and its temperature in an
  analogue black hole}},
  \href{http://dx.doi.org/10.1038/s41586-019-1241-0}{\emph{Nature} {\bf 569}
  (2019) 688--691}, [\href{http://arxiv.org/abs/1809.00913}{{\tt 1809.00913}}].

\bibitem{Isoard:2019buh}
M.~Isoard and N.~Pavloff, \emph{{Departing from thermality of analogue Hawking
  radiation in a Bose-Einstein condensate}},
  \href{http://dx.doi.org/10.1103/PhysRevLett.124.060401}{\emph{Phys. Rev.
  Lett.} {\bf 124} (2020) 060401}, [\href{http://arxiv.org/abs/1909.02509}{{\tt
  1909.02509}}].

\bibitem{Steinhauer:2014dra}
J.~Steinhauer, \emph{{Observation of self-amplifying Hawking radiation in an
  analog black hole laser}},
  \href{http://dx.doi.org/10.1038/NPHYS3104}{\emph{Nature Phys.} {\bf 10}
  (2014) 864}, [\href{http://arxiv.org/abs/1409.6550}{{\tt 1409.6550}}].

\bibitem{Drori:2018ivu}
J.~Drori, Y.~Rosenberg, D.~Bermudez, Y.~Silberberg and U.~Leonhardt,
  \emph{{Observation of Stimulated Hawking Radiation in an Optical Analogue}},
  \href{http://dx.doi.org/10.1103/PhysRevLett.122.010404}{\emph{Phys. Rev.
  Lett.} {\bf 122} (2019) 010404}, [\href{http://arxiv.org/abs/1808.09244}{{\tt
  1808.09244}}].

\bibitem{Rosenberg:2020jde}
Y.~Rosenberg, \emph{{Optical analogues of black-hole horizons}},
  \href{http://dx.doi.org/10.1098/rsta.2019.0232}{\emph{Phil. Trans. Roy. Soc.
  Lond. A} {\bf 378} (2020) 20190232},
  [\href{http://arxiv.org/abs/2002.04216}{{\tt 2002.04216}}].

\bibitem{Guo:2019tmr}
Y.~Guo and Y.-G. Miao, \emph{{Quasinormal mode and stability of optical black
  holes in moving dielectrics}},
  \href{http://dx.doi.org/10.1103/PhysRevD.101.024048}{\emph{Phys. Rev. D} {\bf
  101} (2020) 024048}, [\href{http://arxiv.org/abs/1911.04479}{{\tt
  1911.04479}}].

\bibitem{Bera:2020doh}
A.~Bera and S.~Ghosh, \emph{{Stimulated Hawking Emission From Electromagnetic
  Analogue Black Hole: Theory and Observation}},
  \href{http://dx.doi.org/10.1103/PhysRevD.101.105012}{\emph{Phys. Rev. D} {\bf
  101} (2020) 105012}, [\href{http://arxiv.org/abs/2001.08467}{{\tt
  2001.08467}}].

\bibitem{Blencowe:2020ygo}
M.~P. Blencowe and H.~Wang, \emph{{Analogue Gravity on a Superconducting
  Chip}}, \href{http://dx.doi.org/10.1098/rsta.2019.0224}{\emph{Phil. Trans.
  Roy. Soc. Lond. A} {\bf 378} (2020) 20190224},
  [\href{http://arxiv.org/abs/2003.00382}{{\tt 2003.00382}}].

\bibitem{Ge:2010wx}
X.-H. Ge and S.-J. Sin, \emph{{Acoustic black holes for relativistic fluids}},
  \href{http://dx.doi.org/10.1007/JHEP06(2010)087}{\emph{JHEP} {\bf 06} (2010)
  087}, [\href{http://arxiv.org/abs/1001.0371}{{\tt 1001.0371}}].

\bibitem{Ge:2010eu}
X.-H. Ge, S.-F. Wu, Y.~Wang, G.-H. Yang and Y.-G. Shen, \emph{{Acoustic black
  holes from supercurrent tunneling}},
  \href{http://dx.doi.org/10.1142/S0218271812500381}{\emph{Int. J. Mod. Phys.
  D} {\bf 21} (2012) 1250038}, [\href{http://arxiv.org/abs/1010.4961}{{\tt
  1010.4961}}].

\bibitem{Anacleto:2010cr}
M.~A. Anacleto, F.~A. Brito and E.~Passos, \emph{{Acoustic Black Holes from
  Abelian Higgs Model with Lorentz Symmetry Breaking}},
  \href{http://dx.doi.org/10.1016/j.physletb.2010.09.045}{\emph{Phys. Lett. B}
  {\bf 694} (2011) 149--157}, [\href{http://arxiv.org/abs/1004.5360}{{\tt
  1004.5360}}].

\bibitem{Anacleto:2011bv}
M.~A. Anacleto, F.~A. Brito and E.~Passos, \emph{{Supersonic Velocities in
  Noncommutative Acoustic Black Holes}},
  \href{http://dx.doi.org/10.1103/PhysRevD.85.025013}{\emph{Phys. Rev. D} {\bf
  85} (2012) 025013}, [\href{http://arxiv.org/abs/1109.6298}{{\tt 1109.6298}}].

\bibitem{Ge:2015uaa}
X.-H. Ge, J.-R. Sun, Y.~Tian, X.-N. Wu and Y.-L. Zhang, \emph{{Holographic
  Interpretation of Acoustic Black Holes}},
  \href{http://dx.doi.org/10.1103/PhysRevD.92.084052}{\emph{Phys. Rev. D} {\bf
  92} (2015) 084052}, [\href{http://arxiv.org/abs/1508.01735}{{\tt
  1508.01735}}].

\bibitem{Ge:2019our}
X.-H. Ge, M.~Nakahara, S.-J. Sin, Y.~Tian and S.-F. Wu, \emph{{Acoustic black
  holes in curved spacetime and the emergence of analogue Minkowski
  spacetime}}, \href{http://dx.doi.org/10.1103/PhysRevD.99.104047}{\emph{Phys.
  Rev. D} {\bf 99} (2019) 104047}, [\href{http://arxiv.org/abs/1902.11126}{{\tt
  1902.11126}}].

\bibitem{Guo:2020blq}
H.~Guo, H.~Liu, X.-M. Kuang and B.~Wang, \emph{{Acoustic black hole in
  Schwarzschild spacetime: quasi-normal modes, analogous Hawking radiation and
  shadows}}, \href{http://dx.doi.org/10.1103/PhysRevD.102.124019}{\emph{Phys.
  Rev. D} {\bf 102} (2020) 124019},
  [\href{http://arxiv.org/abs/2007.04197}{{\tt 2007.04197}}].

\bibitem{Cunha:2018acu}
P.~V.~P. Cunha and C.~A.~R. Herdeiro, \emph{{Shadows and strong gravitational
  lensing: a brief review}},
  \href{http://dx.doi.org/10.1007/s10714-018-2361-9}{\emph{Gen. Rel. Grav.}
  {\bf 50} (2018) 42}, [\href{http://arxiv.org/abs/1801.00860}{{\tt
  1801.00860}}].

\bibitem{Perlick:2021aok}
V.~Perlick and O.~Y. Tsupko, \emph{{Calculating black hole shadows: review of
  analytical studies}},  \href{http://arxiv.org/abs/2105.07101}{{\tt
  2105.07101}}.

\bibitem{Zakharov:2011zz}
A.~F. Zakharov, F.~De~Paolis, G.~Ingrosso and A.~A. Nucita, \emph{{Shadows as a
  tool to evaluate black hole parameters and a dimension of spacetime}},
  \href{http://dx.doi.org/10.1016/j.newar.2011.09.002}{\emph{New Astron. Rev.}
  {\bf 56} (2012) 64--73}.

\bibitem{Zakharov:2014lqa}
A.~F. Zakharov, \emph{{Constraints on a charge in the Reissner-Nordstr\"om
  metric for the black hole at the Galactic Center}},
  \href{http://dx.doi.org/10.1103/PhysRevD.90.062007}{\emph{Phys. Rev. D} {\bf
  90} (2014) 062007}, [\href{http://arxiv.org/abs/1407.7457}{{\tt 1407.7457}}].

\bibitem{Perlick:2015vta}
V.~Perlick, O.~Y. Tsupko and G.~S. Bisnovatyi-Kogan, \emph{{Influence of a
  plasma on the shadow of a spherically symmetric black hole}},
  \href{http://dx.doi.org/10.1103/PhysRevD.92.104031}{\emph{Phys. Rev. D} {\bf
  92} (2015) 104031}, [\href{http://arxiv.org/abs/1507.04217}{{\tt
  1507.04217}}].

\bibitem{Konoplya:2019hlu}
R.~Konoplya, A.~Zhidenko and A.~Zinhailo, \emph{{Higher order WKB formula for
  quasinormal modes and grey-body factors: recipes for quick and accurate
  calculations}},
  \href{http://dx.doi.org/10.1088/1361-6382/ab2e25}{\emph{Class.\ Quant.\
  Grav.} {\bf 36} (2019) 155002}, [\href{http://arxiv.org/abs/1904.10333}{{\tt
  1904.10333}}].

\bibitem{Cho:2009cj}
H.~Cho, A.~Cornell, J.~Doukas and W.~Naylor, \emph{{Black hole quasinormal
  modes using the asymptotic iteration method}},
  \href{http://dx.doi.org/10.1088/0264-9381/27/15/155004}{\emph{Class. Quant.
  Grav.} {\bf 27} (2010) 155004}, [\href{http://arxiv.org/abs/0912.2740}{{\tt
  0912.2740}}].

\bibitem{Cardoso:2008bp}
V.~Cardoso, A.~S. Miranda, E.~Berti, H.~Witek and V.~T. Zanchin,
  \emph{{Geodesic stability, Lyapunov exponents and quasinormal modes}},
  \href{http://dx.doi.org/10.1103/PhysRevD.79.064016}{\emph{Phys. Rev. D} {\bf
  79} (2009) 064016}, [\href{http://arxiv.org/abs/0812.1806}{{\tt 0812.1806}}].

\bibitem{Konoplya:2002wt}
R.~A. Konoplya, \emph{{Massive charged scalar field in a Reissner-Nordstrom
  black hole background: Quasinormal ringing}},
  \href{http://dx.doi.org/10.1016/S0370-2693(02)02974-X}{\emph{Phys. Lett. B}
  {\bf 550} (2002) 117--120}, [\href{http://arxiv.org/abs/gr-qc/0210105}{{\tt
  gr-qc/0210105}}].

\bibitem{Richartz:2014jla}
M.~Richartz and D.~Giugno, \emph{{Quasinormal modes of charged fields around a
  Reissner-Nordstr\"om black hole}},
  \href{http://dx.doi.org/10.1103/PhysRevD.90.124011}{\emph{Phys. Rev. D} {\bf
  90} (2014) 124011}, [\href{http://arxiv.org/abs/1409.7440}{{\tt 1409.7440}}].

\bibitem{Cho:2011sf}
H.~T. Cho, A.~S. Cornell, J.~Doukas, T.~R. Huang and W.~Naylor, \emph{{A New
  Approach to Black Hole Quasinormal Modes: A Review of the Asymptotic
  Iteration Method}}, \href{http://dx.doi.org/10.1155/2012/281705}{\emph{Adv.
  Math. Phys.} {\bf 2012} (2012) 281705},
  [\href{http://arxiv.org/abs/1111.5024}{{\tt 1111.5024}}].

\bibitem{Konoplya:2017wot}
R.~A. Konoplya and Z.~Stuchl\'\i{}k, \emph{{Are eikonal quasinormal modes
  linked to the unstable circular null geodesics?}},
  \href{http://dx.doi.org/10.1016/j.physletb.2017.06.015}{\emph{Phys. Lett. B}
  {\bf 771} (2017) 597--602}, [\href{http://arxiv.org/abs/1705.05928}{{\tt
  1705.05928}}].

\bibitem{Jusufi:2019ltj}
K.~Jusufi, \emph{{Quasinormal Modes of Black Holes Surrounded by Dark Matter
  and Their Connection with the Shadow Radius}},
  \href{http://dx.doi.org/10.1103/PhysRevD.101.084055}{\emph{Phys. Rev. D} {\bf
  101} (2020) 084055}, [\href{http://arxiv.org/abs/1912.13320}{{\tt
  1912.13320}}].

\bibitem{Cuadros-Melgar:2020kqn}
B.~Cuadros-Melgar, R.~D.~B. Fontana and J.~de~Oliveira, \emph{{Analytical
  correspondence between shadow radius and black hole quasinormal
  frequencies}},
  \href{http://dx.doi.org/10.1016/j.physletb.2020.135966}{\emph{Phys. Lett. B}
  {\bf 811} (2020) 135966}, [\href{http://arxiv.org/abs/2005.09761}{{\tt
  2005.09761}}].

\bibitem{Cai:2020igv}
X.-C. Cai and Y.-G. Miao, \emph{{High-dimensional Schwarzschild black holes in
  scalar-tensor-vector gravity theory}},
  \href{http://arxiv.org/abs/2011.05542}{{\tt 2011.05542}}.

\bibitem{Li:2021zct}
P.-C. Li, T.-C. Lee, M.~Guo and B.~Chen, \emph{{Correspondence of eikonal
  quasinormal modes and unstable fundamental photon orbits for Kerr-Newman
  black hole}},  \href{http://arxiv.org/abs/2105.14268}{{\tt 2105.14268}}.

\bibitem{Iyer:1986np}
S.~Iyer and C.~M. Will, \emph{{Black Hole Normal Modes: A {WKB} Approach. 1.
  Foundations and Application of a Higher Order {WKB} Analysis of Potential
  Barrier Scattering}},
  \href{http://dx.doi.org/10.1103/PhysRevD.35.3621}{\emph{Phys. Rev. D} {\bf
  35} (1987) 3621}.

\bibitem{Schutz:1985km}
B.~F. Schutz and C.~M. Will, \emph{{BLACK HOLE NORMAL MODES: A SEMIANALYTIC
  APPROACH}}, \href{http://dx.doi.org/10.1086/184453}{\emph{Astrophys. J.
  Lett.} {\bf 291} (1985) L33--L36}.

\bibitem{Konoplya:2003ii}
R.~Konoplya, \emph{{Quasinormal behavior of the d-dimensional Schwarzschild
  black hole and higher order WKB approach}},
  \href{http://dx.doi.org/10.1103/PhysRevD.68.024018}{\emph{Phys. Rev. D} {\bf
  68} (2003) 024018}, [\href{http://arxiv.org/abs/gr-qc/0303052}{{\tt
  gr-qc/0303052}}].

\bibitem{Vieira:2021xqw}
H.~S. Vieira and K.~D. Kokkotas, \emph{{Quasibound states of Schwarzschild
  acoustic black holes}},  \href{http://arxiv.org/abs/2104.03938}{{\tt
  2104.03938}}.

\bibitem{Sun:2017eph}
C.~Yu and J.-R. Sun, \emph{{Note on acoustic black holes from black D3-brane}},
  \href{http://dx.doi.org/10.1142/S0218271819500950}{\emph{Int. J. Mod. Phys.
  D} {\bf 28} (2019) 1950095}, [\href{http://arxiv.org/abs/1712.04137}{{\tt
  1712.04137}}].

\bibitem{Hartnoll:2008vx}
S.~A. Hartnoll, C.~P. Herzog and G.~T. Horowitz, \emph{{Building a Holographic
  Superconductor}},
  \href{http://dx.doi.org/10.1103/PhysRevLett.101.031601}{\emph{Phys. Rev.
  Lett.} {\bf 101} (2008) 031601}, [\href{http://arxiv.org/abs/0803.3295}{{\tt
  0803.3295}}].

\bibitem{Domenech:2010nf}
O.~Domenech, M.~Montull, A.~Pomarol, A.~Salvio and P.~J. Silva, \emph{{Emergent
  Gauge Fields in Holographic Superconductors}},
  \href{http://dx.doi.org/10.1007/JHEP08(2010)033}{\emph{JHEP} {\bf 08} (2010)
  033}, [\href{http://arxiv.org/abs/1005.1776}{{\tt 1005.1776}}].

\bibitem{Montull:2011im}
M.~Montull, O.~Pujolas, A.~Salvio and P.~J. Silva, \emph{{Flux Periodicities
  and Quantum Hair on Holographic Superconductors}},
  \href{http://dx.doi.org/10.1103/PhysRevLett.107.181601}{\emph{Phys. Rev.
  Lett.} {\bf 107} (2011) 181601}, [\href{http://arxiv.org/abs/1105.5392}{{\tt
  1105.5392}}].

\bibitem{Salvio:2012at}
A.~Salvio, \emph{{Holographic Superfluids and Superconductors in
  Dilaton-Gravity}},
  \href{http://dx.doi.org/10.1007/JHEP09(2012)134}{\emph{JHEP} {\bf 09} (2012)
  134}, [\href{http://arxiv.org/abs/1207.3800}{{\tt 1207.3800}}].

\bibitem{Bao:2013fda}
N.~Bao, S.~Harrison, S.~Kachru and S.~Sachdev, \emph{{Vortex Lattices and
  Crystalline Geometries}},
  \href{http://dx.doi.org/10.1103/PhysRevD.88.026002}{\emph{Phys. Rev. D} {\bf
  88} (2013) 026002}, [\href{http://arxiv.org/abs/1303.4390}{{\tt 1303.4390}}].

\bibitem{Guo:2014wca}
H.~Guo, F.-W. Shu, J.-H. Chen, H.~Li and Z.~Yu, \emph{{A holographic model of
  d-wave superconductor vortices with Lifshitz scaling}},
  \href{http://dx.doi.org/10.1142/S0218271816500218}{\emph{Int. J. Mod. Phys.
  D} {\bf 25} (2016) 1650021}, [\href{http://arxiv.org/abs/1410.7020}{{\tt
  1410.7020}}].

\end{thebibliography}\endgroup
\end{document}